\documentclass[11pt]{article}
\pdfoutput=1
\usepackage{amsmath}
\usepackage{amssymb}
\usepackage{graphicx,bbm,mathrsfs}
\usepackage{nicefrac}
\usepackage{bbm}
\usepackage{geometry}       
\usepackage{jheppub}     

\usepackage{dcolumn}   
\usepackage{bm}        
\usepackage{graphicx,mathrsfs}
\usepackage{nicefrac}
\usepackage{multirow}
\usepackage{color}

\def\ie{{\it i.e.}}

\newcommand{\be}{\begin{equation}}  
\newcommand{\ee}{\end{equation}}  
\newcommand{\bea}{\begin{eqnarray}}  
\newcommand{\eea}{\end{eqnarray}}  

\begin{document}

\vspace*{1.2cm}

\begin{center}

\thispagestyle{empty}
{\Large\bf Effective theory for neutral resonances and a statistical dissection of the ATLAS diboson excess}\\[10mm]

\renewcommand{\thefootnote}{\fnsymbol{footnote}}

{\large  Sylvain~Fichet$^{\,a}$, Gero von Gersdorff$^{\,a\,}$ 
\footnote{sylvain.fichet@gmail.com,  gersdorff@gmail.com} }\\[10mm]

\addtocounter{footnote}{-1}

{\it $^a$ ICTP South American Institute for Fundamental Research, Instituto de Fisica Teorica,\\
Sao Paulo State University, Brazil \\
}

\vspace*{12mm}

{  \bf  Abstract }
\end{center}

\noindent 


We classify the complete set of dimension-5 operators relevant for the resonant production of a singlet of spin 0 or 2  linearly coupled to the Standard Model (SM).
 We compute the decay width of such states as a function of the effective couplings, and provide the matching to various well-motivated New Physics scenarios. 
We then investigate the possibility that one of these neutral resonances be at the origin of the excess in diboson production recently reported by the ATLAS collaboration.
We perform a shape analysis of the excess under full consideration of the systematic uncertainties to extract the width $\Gamma_{\rm tot}$ of the hypothetical resonance, finding it to be in the range 26 GeV $<\Gamma_{\rm tot}<$ 144 GeV at 95\% C.L. 
We then point out that the three overlapping selections $WW$, $WZ$, $ZZ$ reported by ATLAS follow a joint trivariate Poisson distribution, which opens the possibility of a thorough likelihood analysis of the event rates. The background systematic uncertainties are also included in our analysis.
We show that the data do not require $W\!Z$ production and could thus in principle be explained by neutral resonances. 
We then use both the information on the width and the cross section, which prove to be highly complementary, to test the effective Lagrangians of singlet resonances. 
Regarding specific models, we find that neither scalars coupled via the Higgs-portal nor the Randall-Sundrum (RS) radion can explain the ATLAS anomaly. The RS  graviton with all matter on the infrared (IR) brane can in principle fit the observed excess, while the RS model with matter propagating in the bulk requires the presence of IR brane kinetic terms for the gauge fields.

\newpage

\section{Introduction}

New particles with TeV masses, neutral under the Standard Model (SM) are a common prediction of various New Physics (NP) scenarios. 
Examples include the Kaluza-Klein (KK) graviton and the radion in warped extra dimensions~\cite{Randall:1999ee}, the dilaton in theories of strongly coupled electroweak breaking \cite{Goldberger:2008zz}, Goldstone bosons of extended composite Higgs models \cite{Kaplan:1983sm}, mesons and glueballs of strongly-coupled theories \cite{Hill:2002ap}, extra scalars breaking the global symmetry of composite Higgs models \cite{vonGersdorff:2015fta}, Higgs  portal models \cite{Schabinger:2005ei},
and  many more.
Among the various SM-singlet resonances, those of spin 2 and spin 0 have strikingly similar couplings to the SM fields, and it is tempting to treat them in a common framework. 

Recently, the ATLAS collaboration has presented a search for narrow resonances decaying to electroweak bosons with hadronic final states using the $8$ TeV LHC dataset~\cite{ATLAS_note}. The weak bosons are highly boosted and are thus reconstructed as a single jet each. A moderate but intriguing excess has been observed near the dijet mass $m_{jj}= $ 2 TeV. It is thus an interesting question whether the diboson excess could be explained by neutral resonances as those predicted in the above scenarios.



The goal of this work is thus to present a unified approach for spin-0 and spin-2 resonances coupled to the SM, and apply it to the search performed in Ref.~\cite{ATLAS_note}.
In a first part, we develop a complete effective field theory (EFT) for neutral resonances of spin 0 and 2.
This general analysis is contained in Sec.~\ref{se:eft}.
As it turns out this EFT consists of only  few operators, which can further be restricted by theoretically well-motivated assumptions, such as approximate flavor and CP conservation. 
All the different neutral resonances listed above then have a simple common description in terms of this effective theory.
Explicit examples of some of these new physics scenarios are then presented  and matched to the EFT Lagrangian in Sec.~\ref{se:scenarios}. Given the concise description of a large class of models in terms of few parameters, our EFT can serve as a model-independent framework that can be applied to any search for resonances at the LHC.

In a second part we then perform a detailed statistical analysis of the ATLAS excess. A basic characterisation of the diboson excess is performed in Sec.~\ref{se:stat}. Local discovery significances are computed in both frequentist and Bayesian frameworks, showing a moderate evidence for the existence of a signal. The shape of the excess is then analysed, taking into account all systematic uncertainties. The total width of the hypothetical resonance is found to be 26 GeV $<\Gamma_{\rm tot}<$ 144 GeV at 95\% C.L. Section \ref{se:statanalysis} contains a comprehensive analysis of the total production rates of the excess. The conditional probabilities for tagging  a true $W$, $Z$ and QCD jet as either $W$ or $Z$ are obtained from the ATLAS simulations, and provide the tagging probabilities for the $WW$, $WZ$, $ZZ$ selections reported by ATLAS. 
 We further observe that these three overlapping selections follow a joint trivariate Poisson distribution, which opens the possibility of a thorough likelihood analysis of the event rates.
The tagging probabilities are checked against the full dataset. 
The estimation of dijet background is treated in a way such that the correlations among the three selections are taken into account. The uncertainty on this background estimation  is then  included as a systematic in the total likelihood. 
Using an actual hypothesis testing, we show that the data do not require $W\!Z$ production and could thus in principle be explained by neutral resonances. 

Finally, in a third part, Sec.~\ref{se:results}, we test the effective Lagrangians of neutral resonance using both the information from the width and from the cross sections. 
It turns out that these pieces of information imply stringent contraints on the EFT parameter space once put together, even after including the uncertainty from the background. These exclusion bounds further imply that various popular scenarios appear to be totally incompatible with the ATLAS diboson excess.

One should remark that various possible scenarios giving rise to the observed ATLAS excess have been considered so far. 
While spin-1 resonances have been investigated by many authors \cite{Fukano:2015hga,Dobrescu:2015qna,Cheung:2015nha,Franzosi:2015zra, Hisano:2015gna,Brehmer:2015cia,Gao:2015irw,Thamm:2015csa,Cao:2015lia,Carmona:2015xaa,Abe:2015uaa,Cacciapaglia:2015eea,Dobrescu:2015yba,Allanach:2015hba,Abe:2015jra,Bian:2015ota,Low:2015uha,Lane:2015fza,Faraggi:2015iaa,Dev:2015pga},
spin-0 and 2 SM-singlet resonances have received far less attention, see~Refs.~\cite{Chiang:2015lqa,Cacciapaglia:2015nga,Sanz:2015zha,Kim:2015vba}. Here we go beyond previous studies by setting up the complete effective theory for neutral resonances. 
We also perform a full statistical analysis of the ATLAS search, 
including the extraction of the width from the shape of the excess.



\section{Effective Field Theory for neutral resonances}

\label{se:eft}

In this section we introduce the EFT of SM-singlet resonances of spin 0 (CP even and odd) and spin 2 coupled linearly to the SM. 
We denote the mass of the resonance with $m$ and assume that it is much heavier than the electroweak (EW) scale, $m^2\gg v^2,\ m_Z^2,\ m_h^2,\, m_t^2$ etc, which is an excellent approximation for a hypothetical 2 TeV resonance. 

We will use field redefinitions (or, equivalently, equations of motion) to reduce the number of independent operators.
The leading interactions will be dimension-5 operators and we denote them generically by $\mathcal O_X$ with coefficients $f_X^{-1}$ where the $f_X$ have dimension of mass. 
The region of validity of the EFT is set by the condition that one can neglect higher dimensional operators. 
The most severe restrictions come from operators with additional derivatives on $\phi$, such as $\partial^2\phi\, G_{\mu\nu}^2$, which require us to impose the condition
\be
m<M\,.
\label{eq:eftregion}
\ee
where $M$ denotes the cutoff of the theory, at which the nonrenormalizable dimension-5 operators become resolved by new states of mass $M$.

In order to estimate the maximal size of the couplings $f_X^{-1}$, we can use Naive Dimensional Analysis (NDA) which gives
\be
f_X^{-1}\lesssim \frac{4\pi }{M}\,.
\label{NDA}
\ee
 Using Eq.~\eqref{eq:eftregion}, the maximal allowed size $f_X^{-1}$ is at most of the inverse EW scale for  $m\sim 2$ TeV.
In many UV completions, the coupling is expected to be weaker than the bound (\ref{NDA}).
For instance, if the nonrenormalizable coupling
\be
\mathcal L_{\phi GG}=f_G^{-1}\phi\, G_{\mu\nu}^2\,,
\label{OGG}
\ee
is resolved in the UV by a heavy fermion of mass $M$, then one expects
\be
f_G^{-1} \lesssim \frac{\alpha_s }{M}\,,
\ee
where $\alpha_s$ is the strong coupling at the scale $M$, and the estimate is obtained by taking the  coupling of $\phi$ to the fermions $\lesssim 4\pi$.


We now list the complete EFT's for the cases of spin 0 (CP odd and even) and spin 2.

\subsection{Spin-0, CP-even}
The effective Lagrangian for a neutral, CP even, spin-0 resonance reads
\be
\mathcal L_{0^+}=
\phi\biggl(f_G^{-1}\, (G^a_{\mu\nu})^2+f_W^{-1}\, (W^i_{\mu\nu})^2+f_B^{-1}\, (B_{\mu\nu})^2
+f_H^{-1}\, |D_\mu H|^2 + 
f_T^{-1}\operatorname{Re}(-y_t\tilde H\,\bar t_Rq_L)\biggr)\,,
\label{scalar}
\ee
where $\tilde H=i\sigma_2 H$.
In order to avoid issues with flavor violation, the operators including fermions are expected to be roughly proportional to the SM Yukawa couplings,  hence here we show only the one involving the top quark, denoted by $q_L$ and $t_R$. 

A priori, one could have written two more operators (that are also relevant for diboson production at the LHC):
\be
\mathcal O_H'=\phi\,\partial^2 |H|^2 \,,\qquad \mathcal O_H''=\phi\, |H|^2\,.
\ee
The operator $\mathcal O_H''$ generates a mass mixing after EWSB as well as a tadpole for $\phi$ that induces a vacuum expectation value (VEV) for this field (or shifts an existing one). It can be eliminated by a field redefinition of $\phi$ in favor of $\mathcal O_H'$, which leaves only kinetic mixing. The operator $\mathcal O_H'$ in turn can be eliminated via the Higgs equations of motion  in favor of $\mathcal O_H$ and $\mathcal O_T$. 
The resulting Lagrangian (\ref{scalar}) does neither have mass nor kinetic mixing between $\phi$ and the Higgs, nor does it induce any VEV for $\phi$.
We will see an expicit example in Sec.~\ref{portal}.
The operator $\phi |H|^4$ gives similar effects as the operator $\phi |H|^2$, but suppressed by an additional factor of $v^2/m^2$.

The decay width resulting from the above Lagrangian (\ref{scalar}) is then~\footnote{We use that $m_W^2,\ m_Z^2,\ m_h^2,\, m_t^2\ll m^2$. The partial decay width to top quarks is suppressed by a relative factor of $m_t^2/m^2$, see below.}
\be
\Gamma=\frac{m^3}{4\pi}\left(\frac{8}{f_G^{2}}+\frac{3}{f_W^2}+\frac{1}{f_B^2}+\frac{1}{8f_H^2}\right)\,.
\ee
The partial widths can easily be extracted from $\Gamma$, see App.~\ref{app:partial}.

A brief comment about the operator $\mathcal O_T$ is in order. The latter can generate couplings of $\phi$ to gluons and photons at one-loop. Even though these cannot be written as local operators, for our purposes (\ie, on-shell production) we can represent this diagram by a complex contribution to e.g.~the $\phi GG$ coupling~\footnote{A similar expression can be given for the $\phi\gamma\gamma$ and $\phi\gamma Z$ couplings which also receive contributions proportional to $f_H^{-1}$ from the $W$-loop.}
\be
\Delta (f_G^{-1})\approx f_T^{-1} \alpha_s (0.0014- 0.0044\, i)\,,
\label{toploop}
\ee 
where we have taken $m=2$ TeV. Eq.~(\ref{toploop}) can simply be obtained from the corresponding expressions of the Higgs couplings to gluons, see e.g.~\cite{Plehn:2009nd,Dumont:2013wma}; note the presence of the imaginary part due to the $t\bar t$-mass threshold. 
Using NDA, $4\pi f_T\gtrsim m$, we obtain the estimate
$|\Delta f_G^{-1}|\lesssim$ (430 TeV)$^{-1}$, and we can safely neglect this contribution to the production of $\phi$.
Moreover, 
$\mathcal O_T$ can induce decays to top quarks with partial width
$\Gamma_{t\bar t}=\frac{3\,m_t^2 m}{32\pi f_T^2}$, which is suppressed by a power of $m_t^2/m^2$ compared to the other decay channels. Only if $f_T^{-1}$ is much larger than all the other couplings will this contribution matter. An upper bound can again be derived using NDA, one finds that $\Gamma_{t\bar t}\lesssim 70$ GeV for $m=2$ TeV.

\subsection{Spin-0, CP-odd}
The Lagrangian for a CP odd, spin-0 resonance is
\be
\mathcal L_{0^-}=
\phi\left( f_G^{-1}\, G^a_{\mu\nu}\tilde G^a_{\mu\nu}
+f_W^{-1}\, W^i_{\mu\nu}\tilde W^i_{\mu\nu}+f_B^{-1}\, B_{\mu\nu}\tilde B_{\mu\nu}+f_T^{-1}\operatorname{Im}(y_t\tilde H \bar t_R q_L)  \right)\,.
\ee
where $\tilde F_{\mu\nu}=\frac{1}{2}\epsilon_{\mu\nu\rho\sigma}F^{\rho\sigma}$.
The additional operators 
\be
\mathcal O_H=\phi\,\partial_\mu \,i[H^\dagger D_\mu H-D_\mu H^\dagger H]
\,,\qquad \mathcal O_{\psi}=\phi\,\partial_\mu (\bar\psi\gamma^\mu \psi) \,,
\ee
where $\psi$ runs over the chiral SM fermions ($\psi=u^i_R$, $d_R^i$, $e_R^i$, $\ell_L^i$ and $q_L^i$),
can all be eliminated by appropriate field redefinitions \footnote{Without loss of generality we have assumed the fermion operators to be flavour-diagonal, though not necessarily flavor-universal (e.g.~$f_{t_R}\neq f_{u_R}$). It should be kept in mind that the degree of flavor-nonuniversality is highly constrained by data. In any case we take only the top-Yukawa coupling to be nonzero. In making field redefinitions of the chiral fermions, one should keep track of anomalies which will generate contributions also to the coefficients of $\mathcal O_{G,W,B}$.} in favor of $\mathcal O_T$. 
Models giving rise to this effective theory have recently been considered in Ref.~\cite{Cacciapaglia:2015nga} in the context of the ATLAS results.
Notice that for $\mathcal O_T$, the same comments as in the CP-even case apply.

The decay width is given by 
\be
\Gamma=\frac{m^3}{4\pi}\left(\frac{8}{f_G^{2}}+\frac{3}{f_W^2}+\frac{1}{f_B^2}\right)\,,
\ee
which is identical to the CP even case, except for the absence of the operator $\mathcal O_H$. Notice that our results agree with those of Ref.~\cite{Kim:2015vba} whereas w.r.t.~Ref.~\cite{Cacciapaglia:2015nga} we find a discrepancy of a factor of 4.
The partial widht to top quarks is again given by $\Gamma_{t\bar t}=\frac{3\,m_t^2 m}{32\pi f_T^2}$ and can be neglected.

\subsection{Spin-2}

We now give the effective Lagragian for CP-even spin-2 fields. 
A massive spin-2 resonance is described by a symmetric-traceless (ST) field $\phi_{\mu\nu}$. As is well known \cite{Fierz:1939ix,Singh:1974qz},
a consistent description requires in addition a scalar field (denoted here by $\chi$), which enforces transversality and removes the unphysical longitudinal degrees of freedom, \ie sets $\partial_\mu \phi_{\mu\nu}=0$, such that only the five physical polarizations remain. Its equation of motion are algebraic, \ie~$\chi$ is a non-propagating auxiliary field (in the absence of sources it simply vanishes, $\chi=0$). We do not write the free Lagrangian here (see however Sec.~\ref{warped}) but rather directly give the propagator, which in the basis $(\phi_{\mu\nu},\chi)$ reads\footnote{Typically the propagator of the massive spin-2 case is given for the reducible representation $\phi_{\mu\nu}+\eta_{\mu\nu}\chi$, see e.g.~\cite{Hinterbichler:2011tt}. Here we prefer to display explicitly the decomposition into the irreducible components.}
\be
P=
i \begin{pmatrix}
\frac{1}{k^2-m^2}\,\Pi^{\mu\nu}_{\rho\sigma}  &
\frac{1}{2m^4}k^{\{\mu}k^{\nu\}}
 \\
\frac{1}{2m^4}k_{\{\rho}k_{\sigma\}}
& \frac{3}{8m^4}(k^2+2m^2)
\end{pmatrix}\label{eq:prop}
\ee
with 
\be
\Pi^{\mu\nu}_{\rho\sigma}=\delta^{\{\mu}_{\{\rho}\delta^{\nu\}}_{\sigma\}}-\frac{2}{m^2}
\delta^{\{\mu}_{\{\rho}k_{\phantom{\{a}}^{\nu\}}k^{\phantom{a\}}}_{\sigma\}}
+\frac{2}{3m^4}
k^{\{\mu}k^{\nu\}}k_{\{\rho}k_{\sigma\}}
\label{projector}
\ee
where the curly brackets denote ST, \ie $X_{\{\mu\nu\}}\equiv \frac{1}{2}X_{\mu\nu}+\frac{1}{2}X_{\nu\mu}-\frac{1}{4}\eta_{\mu\nu}X^\rho_\rho$.
In particular, $P$ mixes the scalar and tensor degrees of freedom, but only the tensor degrees of freedom have physical poles. Notice that on-shell $\Pi$ is simply the projector on transverse, symmetric, traceless fields, in particular~at $k^2=m^2$ one has $\Pi^{\mu\nu}_{\rho\sigma}k^\rho=0$ etc. As usual, the projector can be written in terms of polarization tensors, wich for completeness we collect in App.~\ref{polarizations}.   
The following three observations will further simplify the analysis. 
\begin{enumerate}
\item
As we are only interested in amplitudes for processes near $k^2=m^2$, only the tensor-tensor part of the propagator will matter. In particular, the source of the field $\chi$ \--- which is non-zero in general \--- will not contribute. 
\item
As the tensor-tensor propagator above is transverse on-shell, any source that is just a total derivative of the kind $\partial_\mu J_\nu$, (e.g.,~$\partial_\mu\partial_\nu |H^2|$) will not contribute near the pole. 
\item
Any source that is conserved  (such as $F_{\{\mu}^\rho F_{\nu\}\rho}^{\phantom\rho}$) will only receive contributions from the term proportional to the identity $\delta^{\{\mu}_{\{\rho}\delta^{\nu\}}_{\sigma\}}$.  
\end{enumerate}
The most general bosonic effective Lagrangian linear in a ST field $\phi_{\mu\nu}$ up to dim-5 is then simply
\be
\mathcal L^{\rm bos}_{2^+}=
\phi_{\mu\nu}\biggl(f_G^{-1}\, G^a_{\mu\rho} G^a_{\nu\rho}
+ f_W^{-1}\, W^i_{\mu\rho}W^i_{\nu\rho}+ f_B^{-1}\, B_{\mu\rho} B_{\nu\rho}
+ f_H^{-1}\,D_\mu H^\dagger D_\nu H  
\biggr)\,,
\label{spin2even}
\ee
while the most general fermionic source Lagrangian reads
\be
\mathcal L^{\rm fer}_{2^+}=\phi_{\mu\nu}\biggl(
f_{\psi}^{-1}\operatorname{Im} (\bar\psi \gamma_\mu D_\nu \psi)
\biggr)\,,
\label{spin2evenferm}
\ee
where the sum over chiral fermions ($\psi=u^i_R$, $d_R^i$, $e_R^i$, $\ell_L^i$ and $q_L^i$) is understood. 
We remark that unlike in the scalar cases, even the light SM fermions need to be kept, as one cannot eliminate them via their equations of motion.
Without loss of generality we have diagonalized the operators $\mathcal O_\psi$, but in principle allow non-universal couplings $f_\psi\neq f_{\psi'}$. One should keep in mind though that flavor-nonuniversality (e.g.~$f_{d_R}\neq f_{s_R}$) is highly constrained by data.
It is crucial that one uses the above Fierz-Pauli propagator for the computation of the scattering amplitudes arising from Eqns.~(\ref{spin2even}) and (\ref{spin2evenferm}).~\footnote{Another consistent possibility would be to introduce Goldstone fields $\phi$ and $\phi_\mu$ to render the Lagrangian gauge-invariant under linearized general coordinate transformations, and then adopt a gauge in which the propagators simplify. However, we stress that this procedure also fixes the sources for the auxiliary and Goldstone fields, which cannot be ignored in this case. We will make some more comments on this in Sec.~\ref{warped} in the context of a specific example.}

For the decay width resulting from the above Lagrangian we find (we review in App.~\ref{polarizations} the relevant polarization tensors)
\be
\Gamma=\frac{m^3}{80\pi}\left(\frac{8}{f_G^{2}}+\frac{3}{f_W^2}+\frac{1}{f_B^2}+\frac{1}{12 f_H^2}
+\frac{N_\psi}{4f_{\psi}^2}
\right)\,,
\label{totalspin2}
\ee
where $N_\psi$ denotes the gauge-multiplicity of $\psi$ (QCD plus EW), e.g.~$N=6$ for a LH quark doublet.
Our results agree with Refs.~\cite{Fitzpatrick:2007qr,Agashe:2007zd,Lee:2013bua,Kim:2015vba}.
We refer again to App.~\ref{app:partial} for the decomposition of $\Gamma$ into partial widths.

For completeness, we mention that all symmetric-traceless CP-odd sources up to dimension four made from the SM fields are total derivatives of the kind mentioned in point 2 above, and as such do not contribute to resonant production from dimension-5 operators.~\footnote{We do not agree with some of the results of Ref.~\cite{Kim:2015vba}. Their fermionic source $\tilde T_{2\mu\nu}=\operatorname {Im} \bar q\gamma^5\gamma_{(\mu}\partial_{\nu)}q$ is CP even. In fact the CP-odd sources $\operatorname {Re} \bar q\gamma^5\gamma_{(\mu}\partial_{\nu)}q$ and $\operatorname {Re} \bar q\gamma_{(\mu}\partial_{\nu)}q$ are total derivatives.
}
On the other hand,  dimension-7 operators are always suppressed by additional powers of $m_{Z,W}^2/M^2$, giving only very small cross sections and widths.  
We therefore do not include a CP-odd spin-2 particle in our analysis.

\section{Scenarios}
\label{se:scenarios}

The purpose of this section is to give a few well-motivated scenarios for the effective theories described in Sec.~\ref{se:eft}.

\subsection{Higgs portal}
\label{portal}

Consider a neutral scalar field, of mass $m$, interacting through the Higgs via the interaction
\be
\mathcal L=-\mu\, \phi |H|^2\,.
\label{HP}
\ee
The parameter $\mu$ has dimension of mass and might itself be an effective interaction resulting from some renormalizable coupling $g\Phi^2|H|^2$ after $\Phi$ obtains a vacuum expectation value $\Phi=u+\phi$. 

We now make the shift $\phi\to \phi-\frac{\mu}{m^2}\, |H|^2$. After this field redefinition the Lagrangian becomes
\be
\mathcal L=-\frac{\mu}{m_{\phi}^2}\partial_\mu\phi\,\partial_\mu|H|^2+\frac{1}{2}\frac{\mu^2}{m^2}(\partial_\mu|H|^2)^2+\dots
\label{higgsportal2}
\ee
where the ellipsis denotes unobservable modifications of the Higgs potential.
The first term is the effective interaction discussed below Eq.~(\ref{scalar}) with the identification
\be
f_H'=\frac{m^2}{\mu}\,.
\ee
The second term leads to modifications of the Higgs couplings. In order to avoid too-large deviations inconsistent with experiment, we will impose that $\mu v\ll m_{\phi}^2$, which implies that $v\ll f_H'$. 
To arrive at our standard basis, we use the Higgs equations of motion (or, equivalently, make the field redefintion $H\to H+\frac{\mu}{m^2}\phi H$) to find
\be
\frac{\mu}{m^2}\phi\,\partial^2|H|^2=\frac{2\mu}{m^2}\phi\left(|D_\mu H|^2-y_t\operatorname{Re}(\tilde H\,\bar t_Rq_L)+m_H^2|H|^2-2\lambda |H|^4)
\right)\,.
\ee
The last two terms (that results from the Higgs potential $\mathcal V=-m_H^2|H|^2+\lambda|H|^4$) can be neglected, as they are suppressed w.r.t.~to the original interaction (\ref{HP}) by a factor of $m_H^2/m^2$, $|H|^2/m^2$. We then have
\be
f_H=f_T=\frac{m^2}{2\mu}\,.
\ee
with the remaining $f_X^{-1}$ vanishing.


\subsection{Spin-2 Lagrangians from warped extra dimensions.}
\label{warped}

In this section we derive the massive interacting spin-2 Lagrangian from a warped extra dimension \cite{Randall:1999ee}.
According to the general discussion in Sec.~\ref{se:eft}, we expect the presence of an auxiliary field. Moreover, in the extra dimensional construction we arrive naturally at a theory containing Goldstone modes as extra-dimensional components of the metric, which one can simply set to zero in a "unitary" gauge.

We then consider a 5d compactification in the metric background 
\be
d s^2=(kz)^{-2}(d x^2-d z^2)\equiv\gamma_{MN}dx^Mdx^N\,,
\label{ads}
\ee
where $z$ denotes the 5th coordinate $z_0<z<z_1$ and $k=z_0^{-1}$ the Anti-de-Sitter curvature .

After decomposition in Kaluza Klein (KK) modes, the kinetic Lagrangian of the fluctuations of the 5d metric becomes\footnote{We refer the reader to Ref.~\cite{Dudas:2012mv} for details, in particular the precise relation of the various fields to the 5d metric. Eq.~(\ref{L5d}) is obtained from Eq.~(3.5) of \cite{Dudas:2012mv} by use of the 5d wave functions $f_s=\sqrt{2}z^sJ_s(m_n z)/z_1J_2(m_n z_1)$, where the $J_\nu$ denote Bessel functions. The masses are solutions to $J_1(m_nz_1)=0$ and $\phi_{\mu\nu},\ \chi$ have wave functions $f_2$, $\phi_\mu$ has wave function $f_1$, and $\phi$ has wave function $f_0$.}
\begin{multline}
\mathcal L_{\rm kin}=
-\frac{1}{2}\phi^n_{\mu\nu}(\partial^2+m_n^2)\phi^n_{\mu\nu}+\frac{1}{2}\phi^n_{\mu}(\partial^2+m_n^2)\phi^n_{\mu}+\frac{1}{2}\chi^n(\partial^2+m_n^2)\chi^n
-\frac{1}{2}\phi^n(\partial^2+m_n^2)\phi^n\\
-\left(\partial_\mu\phi^n_{\mu\nu}+\frac{1}{2}\partial_\nu\chi^n+\frac{m_n}{\sqrt 2}\phi^n_\nu\right)^2
+\left(\frac{1}{\sqrt{2}}\partial_\mu \phi^n_\mu+m_n\chi^n+\sqrt{\frac{3}{2}}m_n\phi^n\right)^2\,.
\label{L5d}
\end{multline}
Notice that the Lagrangian is completely diagonal in the KK modes.
Here, $\phi_\mu^n$ and $\phi^n$ denote the Goldstone modes originating from the extra-dimensional components of the 5d metric,\footnote{The field $\phi_{\mu}$ does not have a zero mode, while $\phi$ has a zero mode that is not eaten and corresponds to the radion.} and $\chi^n$ is the above mentioned auxiliary field. Setting to zero the Goldstone fields 
\be
\phi^n=0\,,\qquad \phi_\mu^n=0 \,,\qquad n\neq 0\,,
\label{unitary}
\ee
one arrives at the unitary gauge, 
which is precisely the Fierz-Pauli Lagrangian \cite{Fierz:1939ix,Singh:1974qz} leading to the propagator (\ref{eq:prop}). 
Instead, one could adopt the Feynman gauge, in which the terms in the second line of Eq.~(\ref{L5d}) are cancelled by an appropriate Fadeev-Popov procedure.
In Feynman gauge, the propagators are especially simple, in particular, all fields have the same mass and do not mix; observe that the field $\chi$ becomes propagating and has a "wrong sign" kinetic term.
We however stress that in this case, the sources for {\em all} fields, $\phi^n_{\mu\nu}$, $\chi^n,$ $\phi^n_\mu$ and $\phi^n$  have to be taken into account. 
In the following we will employ the unitary gauge (\ref{unitary}) in which case we only need to consider the source for the ST field $\phi_{\mu\nu}^n$.

As for the interactions, typically two scenarios are considered. In the brane model, all SM fields are localized on the IR brane \cite{Randall:1999ee}, while in the bulk model they are allowed to propagate in the bulk \cite{Gherghetta:2000qt,Agashe:2007zd,Huber:2000ie}. In the latter case, the gauge fields have flat 5d profiles, the RH top and the Higgs fields have profiles peaked towards the IR brane, and the remaining matter fields have profiles that are flat or peaked towards the UV brane.\footnote{We remark that such a scenario features other states, typically lighter than the KK graviton, whose phenomenology will severley constrain the model. We will not further consider these model-dependent constraints in this work. } 
For our purposes it is  good enough to approximate the bulk model by IR brane localized RH top and Higgs fields and completely ignore the other quarks and leptons.
The interaction terms for IR-brane localized fields are given by
\be
\mathcal L_{\rm int}^{\rm matter}
=\frac{z_1 k}{M_P}\phi^n_{\mu\nu}\left(2D_\mu H^\dagger D_\nu H
%
-\operatorname{Im}\bar t_R\gamma_\mu D_\nu t_R
\right)\,,
\ee
where $M_P$ is the reduced Planck mass. In the scenario with all SM fields localized on the IR brane, there are identical contributions to the remaining SM fermions.
Gauge fields couple as (for any $n\neq 0$):
\be     
\mathcal L_{\rm int}^{\rm gauge}=
\frac{z_1 k}{M_P}\,\zeta_n\,\phi^n_{\mu\nu}F_{\mu\rho}F_{\rho\nu}\,,\qquad \zeta_n\equiv \frac{r_1+2\,x_n^{-2}
[1+J_2(x_n)^{-1}]
}{r_0+r_1+V}\,,
\ee
where $x_n=z_1 m_n$ and 
the quantities $r_0$ and $r_1$ denote possible brane kinetic terms (BKT) \cite{Fichet:2013ola}, and $V=\log(kz_1)\approx 36$ is the volume of the extra dimension. 
An IR brane-localized gauge field is described by the limit $r_1\to \infty$, or $\zeta_n=1$.\footnote{We remark that in this limit the  KK modes of the gauge fields become strongly coupled $g_{\rm KK}^2\sim g^2(V+r_1)$, hence to avoid the non-perturbative regime one would demand 
$\sqrt{V+r_1}<4\pi/ g$. Note also that the gauge KK modes decouple from the IR brane in this limit.\label{foot:pert}}

For the bulk model, the effective Lagrangian for the first KK mode of mass $m$ is then given by $\mathcal L_{2^+}$ defined in Eqns.~(\ref{spin2even}) and (\ref{spin2evenferm}), with the couplings
\be
f_H=\frac{m}{2\kappa }\,x_1^{-1} \,,\qquad f_G=f_W=f_B=-\frac{m }{\kappa}\,(\zeta_1 x_1)^{-1} \,,\qquad f_{t_R}=-\frac{m}{\kappa }\,x_1^{-1}
\ee
where $x_1=3.83$, and $\kappa=k/M_P$ the RS coupling parameter. 
For the brane model, one has instead
\be
f_H=\frac{m}{2\kappa }\,x_1^{-1} \,,\qquad f_G=f_W=f_B=-\frac{m }{\kappa}\,x_1^{-1}\,,\qquad f_{\psi}=-\frac{m}{\kappa }\,x_1^{-1}\,.
\ee
According to our general formula Eq.~(\ref{totalspin2}), the bulk and brane model's total widths are respectively
\be
\Gamma_{\rm bulk}
=\frac{(13+144\zeta_1^2)\,(\kappa x_1)^2m}{960 \pi}\,,\qquad 
\Gamma_{\rm brane}
=\frac{283(\kappa x_1)^2m}{960\pi}\,.
\ee
In the RS bulk model, the terms proportional to $\zeta_1^2$ contribute $0.2\%$, $26\%$, and $56\%$ to the total width for $r_1/V=0$, $0.2$ and $0.5$ respectively.

\subsection{Radion/Dilaton}

In the warped extra-dimensional scenario considered in the previous section, the field $\phi^0$ corresponds to the radion which  describes the fluctuations of  the size of the extra dimension. 
It is massless in the background (\ref{ads}) but by a suitable stabilization mechanism it acquires a mass \cite{Goldberger:1999uk,DeWolfe:1999cp,Cabrer:2009we}. Athough its five-dimensional wave-function is deformed by the stabilization mechanism\footnote{See Refs.~\cite{Medina:2010mu,Cabrer:2011fb} for some analytic expressions for the stabilized profile.} we will assume that these effects are small and its couplings are thus approximated by those of the massless case. One finds (see e.g.~Ref.~\cite{Csaki:2007ns})  
\be
\mathcal L_{\rm int}=\frac{k z_1}{\sqrt{6}M_P}\phi\left(
\frac{1}{4(V+r_0+r_1)}\,F_{\mu\nu}^2+2\,|D_\mu H|^2-2\operatorname {Re}\, (y_t\tilde H \bar t_R q_L) 
\right)
\ee
There is an additional coupling proportional to the Higgs potential $\mathcal V(H)=-m_{H}^2|H|^2+\lambda |H|^4$. Eliminating the operator $\phi|H|^2$ will result in negligible corrections to the operator coefficients $f_H^{-1},f_{T}^{-1}$ suppressed as $m_H^2/m^2$.

It is customary to treat the radion interaction scale defined as
\be
f_{\rm rad}=\frac{\sqrt{6}M_P}{kz_1}
\ee
as a free paramter. 
In the bulk model one then finds the couplings
\be
f_H=f_T=\frac{f_{\rm rad}}{2}\,,\qquad f_G=f_W=f_B=4 V f_{\rm rad}
\ee
where we have assumed $r_i\ll V$.
The brane model is again obtained by sending $r_1\to\infty$. The couplings to the gauge boson field strength vanishes in this case, and one is left with only
\be
f_H=f_T=\frac{f_{\rm rad}}{2}.
\ee
Interestingly, the brane model effective Lagrangian precisely conincides with the Higgs portal scenario with the identification $f_{\rm rad}=m^2/\mu$.
In either case, the field $\phi$ just inherits the Higgs couplings suppressed by a factor $v/f_{\rm rad}$.
As the couplings to gauge bosons are always small, the decay width comes entirely from $f_H$ in both models
\be
\Gamma=\frac{m^3}{8\pi f_{\rm rad}^2}.
\ee
As explained in Sec.~\ref{scalar}, the decay to tops are suppressed by $m_t^2/m^2$ and do not contribute to the total width.
Finally we recall that the radion is closely related to the dilaton of nearly conformal extensions of the SM, so that very similar results hold in this case.


\section{Characterisation of the ATLAS diboson excess} \label{se:stat}

\subsection{Data, background and local significances} \label{se:data}

The ATLAS collaboration has recently presented a search for narrow resonances decaying to electroweak bosons  with hadronic final states using the $8$ TeV LHC dataset~\cite{ATLAS_note}. This dataset has   $20.3$~fb$^{-1}$ integrated luminosity. 
The weak bosons from massive resonances are highly boosted and are thus reconstructed as a single  jet with large radius using advanced reclustering, grooming and filtering algorithms. 
The expected background is dominated by dijets events from the QCD background, which is huge but does not feature potential resonance structures.

Boson-tagging cuts are applied to the selected dijet events, asking for subjet momentum-balance and low number of associate charged particles tracks. Each jet is then tagged using a narrow window on the jet mass $m_j$, asking for $m_j$ to be close to the $W$ or $Z$ mass. In the analysis, a jet is identified as a $W$ if $m_j\in[69.4,95.4]$~GeV, and is identified as $Z$ if $m_j\in[79.8,105.8]$ GeV. The $W$ and $Z$ masses being close, these two ranges overlap. There are thus three disjoint tagging regions, that we label as $W$-only, $W$ or $Z$ (noted $W/Z$), and $Z$-only.

A local excess of observed events appears in the dijet spectrum near $2$ TeV. The numbers reported in Ref.~\cite{ATLAS_note} (and its  extra material \cite{ATLAS_note_extra}) in the three bins $m_{jj}\in[1850,1950]$, $[1950,2050]$, $[2050,2150]$, that we refer to as the excess region,  are shown in Tab.~\ref{tab:data_bkg}.  The expected dijet background in each bin is also shown. The background is partly determined from a fit to the whole dijet spectrum, and is thus subject to some uncertainty.

\begin{table}
\center
\begin{tabular}{|c|ccc|ccc||c|c|}
\hline & \multicolumn{3}{c}{ $\hat n_r$}  & \multicolumn{3}{|c||}{ $b_r$} & $\hat{n}$ & $b$  \\ \hline
$WW$ & $4$ & $7$ & $2$ & $2.67^{+0.42}_{-0.40}$ & $1.84^{+0.34}_{-0.31}$ & $1.30^{+0.28}_{-0.24}$ & $13$ & $6.61^{+1.22}_{-1.01}$ \\ \hline
$WZ$ & $5$ & $8$ & $2$ & $3.12^{+0.50}_{-0.42} $ & $2.08^{+0.39}_{-0.33} $ & $1.41^{+0.33}_{-0.26}  $  & $15$ & $5.81^{+1.04}_{-0.95}$ \\ \hline
$ZZ$ & $5$ & $3$ & $1$ & $0.91^{+0.23}_{-0.20} $ & $0.55^{+0.16}_{-0.13} $ & $0.34^{+0.12}_{-0.09} $ & $9$ &
$1.8^{+0.50}_{-0.42}$ \\ \hline
\end{tabular}
\caption{
Data  and background obtained from \cite{ATLAS_note} in the three bins $r=\{[1.85, 1.95]$, $[1.95, 2.05]$, $[2.05,2.15]~{\rm TeV}\}$ and in the whole excess region of the dijet mass spectrum,  for the $WW$, $WZ$ and $ZZ$ selections. 
\label{tab:data_bkg}}
\end{table}
As a first step, one should check the statistical significance of this excess. Assuming Poisson statistics for the observed events in each bin, we first compute the  p-value of a discovery test in every bin. 
This computation is done with and without taking into account the background uncertainties, that we model using a nuisance parameter $\theta\in[\theta_a,\theta_b]$ with a flat ``prior'' distribution.  

The likelihood for one of the bins $r$ simply reads
\be
L(s_r,\theta)=\frac{(s_r+b_r+\theta)^{\hat n _r}e^{-s_r-b_r-\theta}}{\hat n_r !}\,.
\ee
The nuisance parameter is eliminated by maximising this likelihood with respect to $\theta$ for a given $s_r$, $\bar L(s_r)=\max_{\theta}L(s_r,\theta)$. 
The statistical significance $Z_0$ for the existence of an excess is obtained by computing the probability density $f_q$ for $q=-2\log[\bar L(s_r)/\max_{s_r}\bar L(s_r)]$ and evaluating the observed p-value $p=\int_{q_{\rm obs}}^\infty dq\,f_q $. The p-value is further translated into a standard significance by $Z_0=\Phi^{-1}(1-p)$, where $\Phi$ is the standard cumulative Gaussian distribution.
One allows for both upward and downward fluctuations.   The significance of this discovery test is computed for each bin. The values, shown in Tab.~\ref{tab:tests},  typically go beyond two sigmas in the central bin.

\begin{table}
\center
\begin{tabular}{|c|ccc|ccc|ccc|}
\hline & \multicolumn{3}{c}{ $Z_0$ (without syst.)}  & \multicolumn{3}{|c|}{ $Z_0$ (with syst.)}  
& \multicolumn{3}{|c|}{ $B_0$ } 
\\ \hline
$WW$ & $<1$ & 2.8 & $<1$ & $<1$ & 2.5 & $<1$ & 0.9 & 13 & 0.8 \\ \hline
$WZ$ & $<1$ & 3.0 &  $<1$ &  $<1$ & 2.7 & $<1$ &  1.0 & 22 & 0.8 \\ \hline
$ZZ$ & 2.8 & 2.1  & $<1$ & 2.6 & 1.8 & $<1$ & 21 & 6 & 1.1 \\ \hline
\end{tabular}
\caption{
Local discovery tests for the $[1.85, 1.95]$, $[1.95, 2.05]$, $[2.05,2.15]$ bins of the 
 dijet mass distribution with $WW$, $WZ$ and $ZZ$ tagging. 
 Left and middle columns: discovery significance without and with systematic uncertainties. Right column: discovery Bayes factor.
\label{tab:tests}}
\end{table}
We also introduce a Bayesian discovery test, defined as
\be
B_0=\frac{p(\hat n_r|s_r)}{p(\hat n_r|s_r=0)}
\ee
This expression takes the simple form
\be
B_0=\frac{\int d s_r\, L(s_r) \pi(s_r)}{L(0)}\,.
\ee
It turns out that the prior for the signal $\pi(s_r)$ is entirely fixed from general considerations. Indeed, the measurement being a counting experiment, we already know a priori that $b_r+s_r$ follows a Poisson distribution. The parameter of this Poisson distribution has to be chosen to be $b_r$, which is  known a priori, in order not to bias the discovery test. This then fixes $\pi(s_r)$ to be \footnote{In \cite{shape_paper_WIP} it will be shown that this particular prior provides a good connexion between discovery Bayes factor and frequentist statistical significance. Also, notice that we do not implement the background systematic error in the Bayes factor. This is because this type of systematic uncertainty approximately cancels out  in the Bayes factor, as will be shown  in \cite{shape_paper_WIP}. } 
\be
\pi(s_r)=\frac{b_r^{b_r+s_r}e^{-b_r}}{(b_r+s_r)!}\,.\label{eq:piPois}
\ee
The values of the discovery Bayes factor  are shown in Tab.~\ref{tab:tests}. One can see that the values of $B_0$ are beyond the threshold of moderate evidence for the central bin. 

It follows that both frequentist and Bayesian discovery tests provide a moderate evidence for the existence of a local excess over the QCD dijet background. We conclude that this excess is significant enough to deserve attention, so that we proceed in the analysis.

\subsection{Mass and width reconstruction}\label{se:stat_shape}

As the data are provided in several bins, it is possible to analyse the shape of the hypothetical signal.
Even though the  statistics of the excess is fairly low, we emphasize that there is no reason that prevents  to apply a rigorous shape analysis. Whether or not the data are informative enough should be decided by the outcome of the analysis.
Notice that, as the excess is observed in more than one bin, one can expect \textit{both} an upper and lower limit on the width of the resonance.

In what follows the bins of the $m_{jj}$ distribution are labelled by the index $r$.
Contrary to the analysis on the total event numbers, here we do not combine the events of the three selections $WW$, $WZ$, $ZZ$, and rather perform the shape analysis for each selection separately. It will be clear from next section that a more evolved analysis combining the three selections would bring only little extra information.

 The likelihood containing  the shape information appears naturally from the full likelihood $L=\prod_r \,L_r$,  by factoring out the likelihood for the total event number, $L=L^{\rm tot}L^{\rm shape}$. Explicitly, the shape likelihood reads 
\be
L^{\rm shape}= \prod_{r~{\rm (bins)}} \left(\frac{n_r}{n_{\rm tot}}\right)^{\hat n_r}\,.
\ee
Note that the factorisation $L=L^{\rm tot}L^{\rm shape}$ makes clear that  a shape analysis of the diboson excess is truely \textit{complementary} from the  total event number analysis, because each analysis rely on mutually exclusive pieces of information.

We denote the shape of the expected signal by a distribution $f_{m_{jj}}$ normalised to one (\ie~a density).
The shape of the signal is modelled assuming a  resonant amplitude, and the background is assumed to be flat near the peak of the resonance. The narrow-width approximation is assumed, \ie~one takes $\Gamma/m \ll 1$, that will be well verified a posteriori. 
Given these standard assumptions, the $m_{jj}$ distribution is then distributed following  a Breit-Weigner shape,
\be
f_{m_{jj}}\propto\frac{1}{(m_{jj}^2-m^2)^2+m^2 \Gamma^2}\,.
\ee
The expected content of the bins is obtained by integrating over this distribution,
\be
n_r =n_{\rm tot}\, \int_{{\rm bin}~r} f_{m_{jj}} \,,
\ee
an one will note $f_r$ the shape density integrated over a bin, $f_r\equiv\int_{{\rm bin}~r} dm_{jj} \,f_{m_{jj}}$.
We consider  the three bins centered around $2$ TeV, and assume no signal event elsewhere.  

We also take into account the  systematic uncertainties relevant for the shape of the signal. 
These are the uncertaintites on the jet reconstruction (see \cite{ATLAS_note}), that tend to smear the resonance shape. The sources of error are the jet $p_T$ resolution, the jet $p_T$ scale and the jet mass determination, associated respectively to the nuisance parameters $\delta_{\rm res}$, $\delta_{\rm scale}$, $\delta_{\rm m}$, affecting the $m_{jj}$ mass. 
 The magnitude of these errors is small with respect to one, so that they can be written in the linear form  \be m_{jj}(1+\delta_{\rm res}+\delta_{\rm scale}+\delta_{\rm m})\,.\ee
All these uncertainties are modelled using Gaussian nuisance parameters $\delta_{\rm res}$, $\delta_{\rm scale}$, $\delta_{\rm m}$ with zero mean and respective standard deviation $\sigma_{\rm res}=0.033$, $\sigma_{\rm scale}=0.02$, $\sigma_{\rm m}=0.03$ (see \cite{ATLAS_note},~Tab.~4).

These three nuisance parameters being independent, they can be rigorously combined into a single Gaussian nuisance  parameter $\delta$ with zero mean and variance given by
\be
\sigma^2=\sigma_{\rm res}^2+\sigma_{\rm scale}^2+\sigma_{\rm m}^2\,.
\ee
 The event number in a given bin depends thus on $m,\Gamma,\delta$, so that the complete likelihood for the shape analysis of the diboson excess reads
\be
L(m,\Gamma)=\int d\delta\,\prod_{r~{\rm (bins)}} \big[f_r(m,\Gamma,\delta)\big]^{\hat n_r} \,\pi(\delta) 
\ee
For the mass and width of the hypothesized resonance, one assumes log priors $\pi(m)\propto m^{-1} $, $\pi(\Gamma)\propto \Gamma^{-1} $, which are the most objective priors for dimensionful quantities.
The confidence regions are drawn from the posterior density, which is given by $p(m,\Gamma)=L(m,\Gamma)\pi(m)\pi(\Gamma)$.  

\begin{figure}
\centering{
\includegraphics[scale=0.36,clip=true, trim= 0cm 0cm 0cm 0cm]{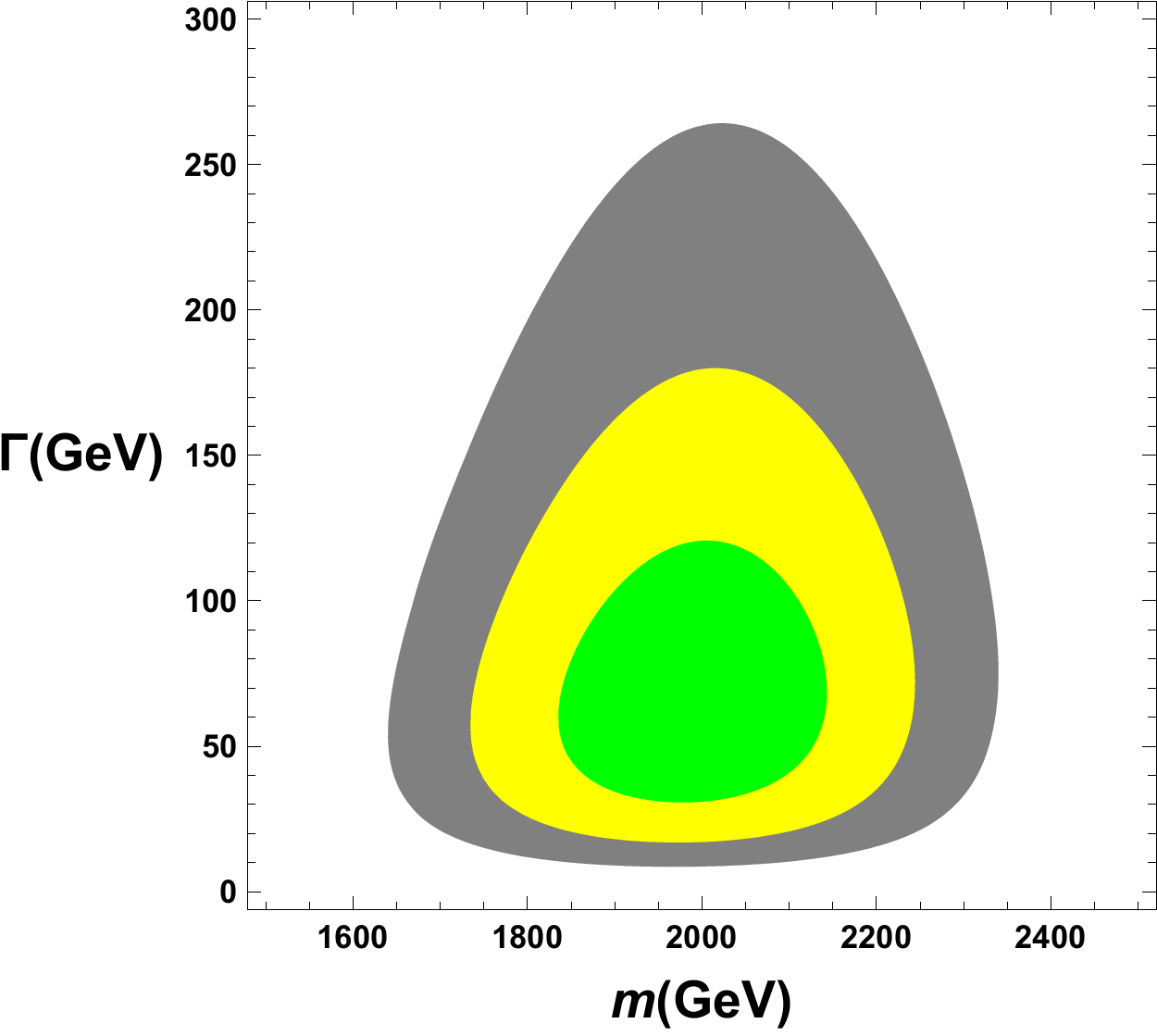}
\quad
\includegraphics[scale=0.36,clip=true, trim= 0cm 0cm 0cm 0cm]{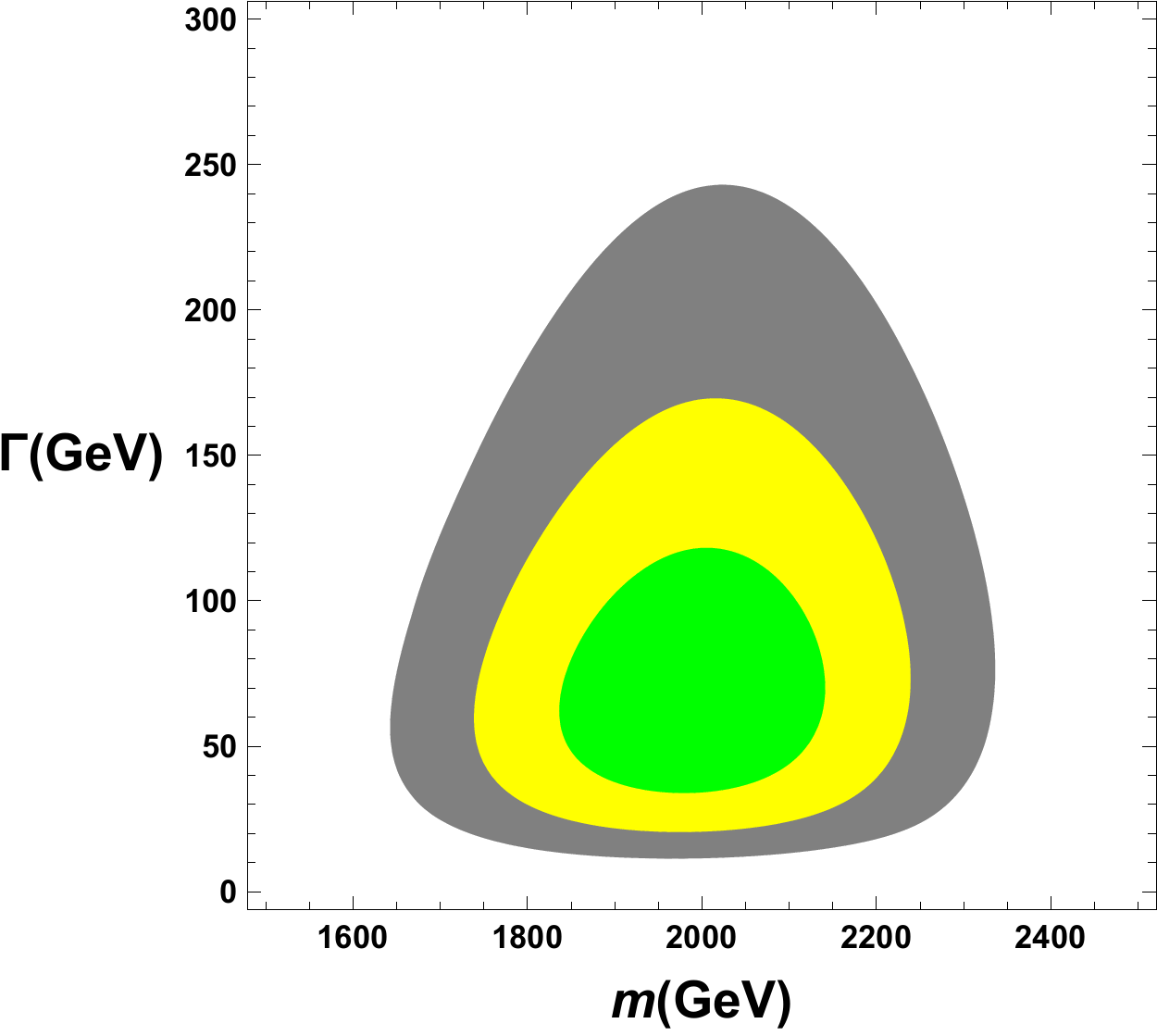}
\quad
\includegraphics[scale=0.36,clip=true, trim= 0cm 0cm 0cm 0cm]{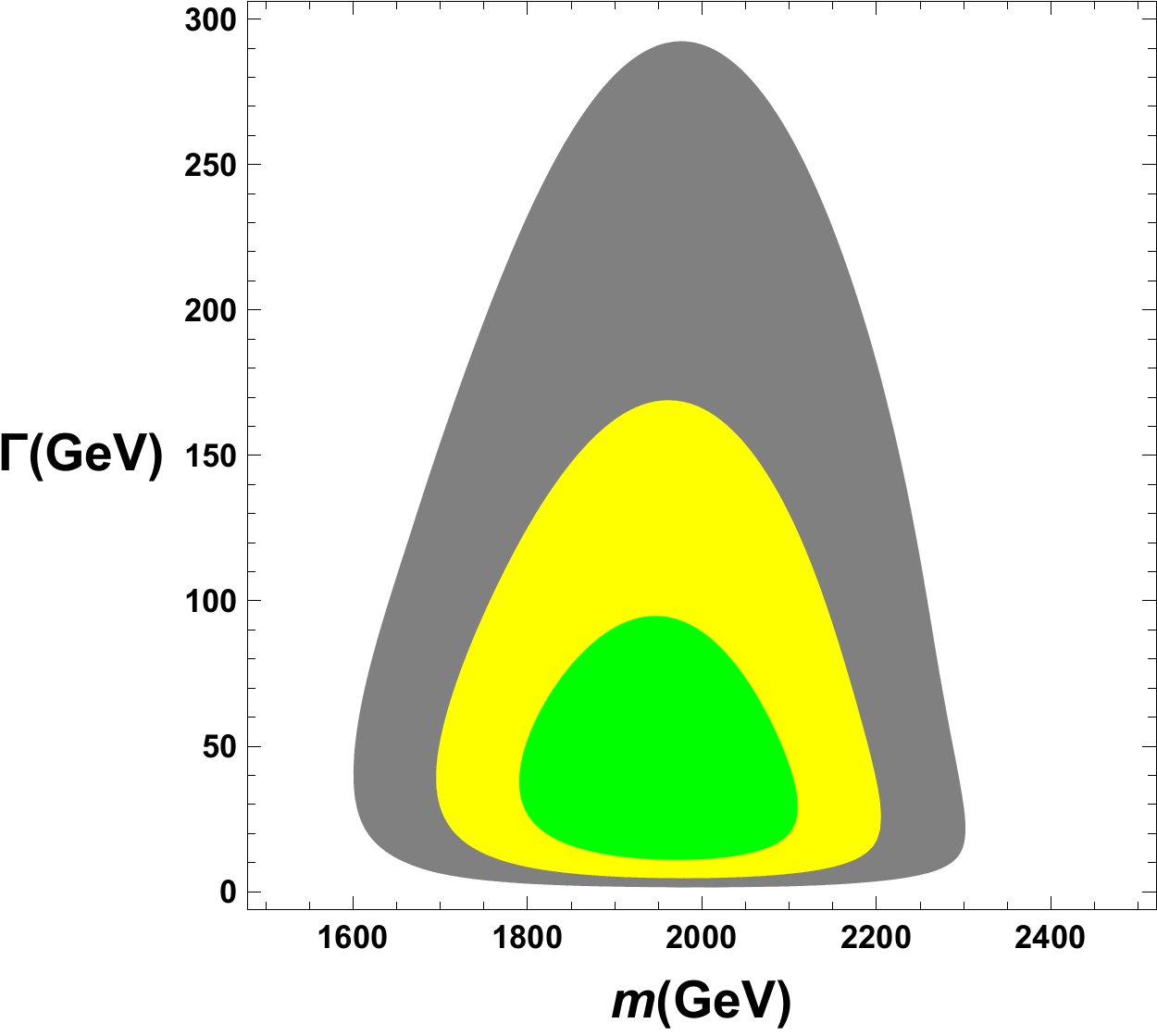}}
\caption{
Probability densities in the $m-\Gamma$ plane, for $WW$, $WZ$ and $ZZ$ selections (from left to right). The green, yellow and gray regions correspond respectively to $68\%$, $95\%$ and $99\%$ confidence level.
\label{fig:2DBCI}}
\end{figure}

The one-dimensional confidence intervals for mass and width are given  in Tab.~\ref{tab:1DBCI}.  The systematics errors increase the mass CL bounds by roughly $\sim 5\%$ and the width CL bounds up to $\sim 20\%$. Using a flat prior instead of a log prior changes the bounds by 
roughly $\sim 10\%$. 
The two dimensional confidence regions in the $m-\Gamma$ plane are shown in Fig.~\ref{fig:2DBCI}. In the following, we shall quote the results from the $WZ$ selection, which contains the largest event number.

\begin{table}
\center
\begin{tabular}{|c|c|c|c|c|}
\hline
& \multicolumn{2}{c}{ Mass [GeV]}  & \multicolumn{2}{|c|}{ Width  [GeV]}  
\\ 
& $68\%$ CL & $95\%$ CL  & $68\%$ CL & $95\%$ CL
\\ \hline
$WW$ & $[1895,2091]$ & $[1797,2190]$ &$[ 39 ,99 ]$ & $[22,150]$  \\ \hline
$WZ$ & $[1895,2091]$ &$[1797,2189]$ &$[42,98]$ &$[26,144]$ \\ \hline
$ZZ$ &$[1852,2047]$ &$[1755,2145]$ &$[15,71]$ &$[6,139]$ \\ \hline
\end{tabular}
\caption{ One-dimensional confidence intervals at $68\%$ and $95\%$ confidence level  for the mass ($m$) and the width ($\Gamma$) of the hypothesized resonance. The intervals are computed independently for each subchannel.
\label{tab:1DBCI}}
\end{table}

\section{Statistical analysis of the diboson rates}
\label{se:statanalysis}


Having studied the shape of the diboson excess, we now turn to the analysis of the overall event numbers, \ie~the total rates over the excess region. 
The likelihood analysis for a set of overlapping selections is a somewhat unusual exercise to carry out, so that we shall provide a detailed explanation of the statistics involved.  
For clarity, in the following we will use rigorous probability notation.
 The hypothetical event number in a given selection is taken as random variable, denoted by $N$. 
 Specific values of event numbers are denoted by $n$, and $P(N=n)$ is the probability of $N$ for taking the value $n$.  The expected event numbers are denoted by $\lambda$, and the observed event numbers are denoted by $\hat n$. 

\subsection{The statistics of hadronic weak-boson tagging}
\begin{figure}
\begin{picture}(400,150)
\put(120,0){\includegraphics[scale=0.6]{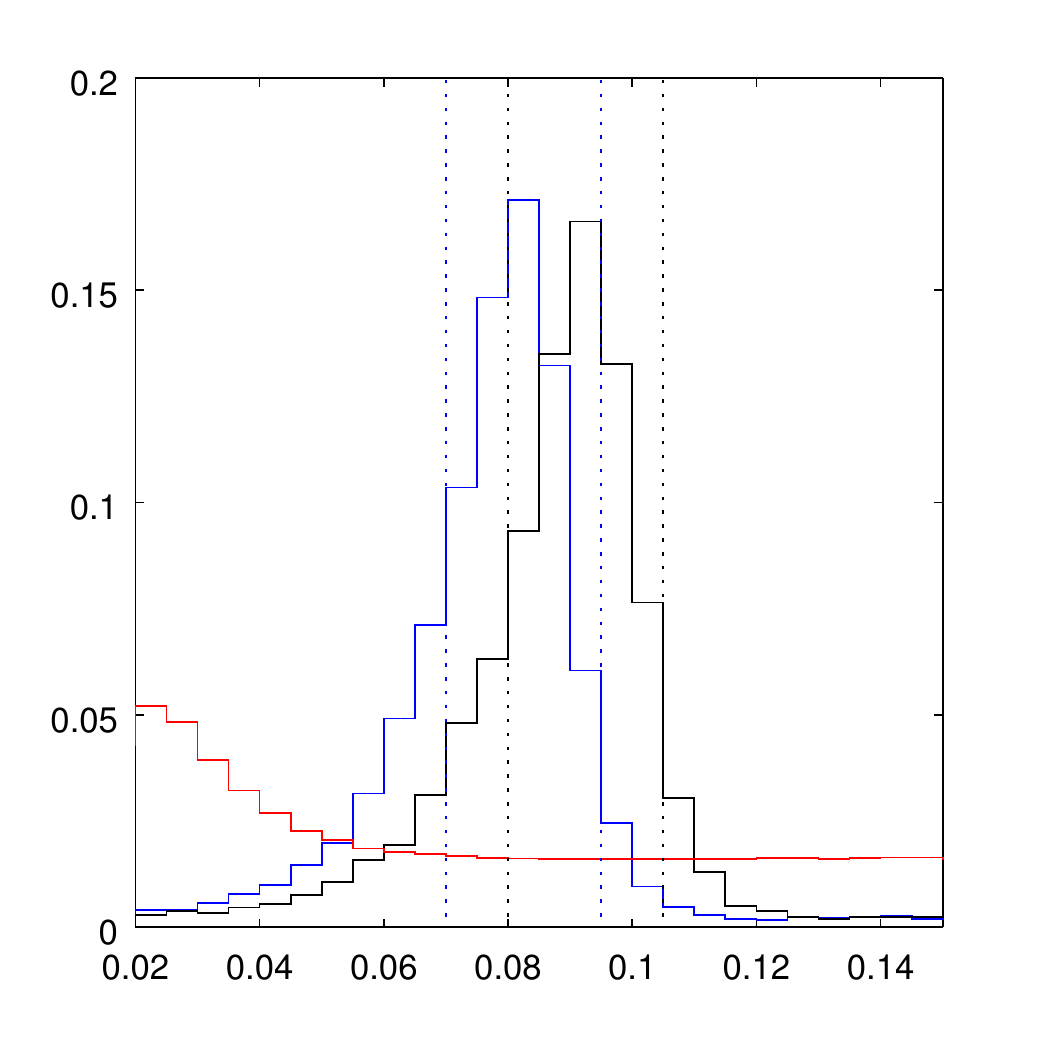}}
\put(200,-4){$m_j$ [TeV]}
\end{picture}
\caption{ Mass distributions of a jet  arising from a $W$ (blue curve), $Z$ (black curve) and the QCD dijet background (red curve). 
Blue and black dotted lines represent the $W$ and $Z$ tagging regions respectively, giving rise to the three disjoint tagging regions, $W$-only, $W/Z$ and $Z$-only.
\label{fig:mj_tag}}
\end{figure}

The  mass distribution of a fat jet coming from a  $W$ or $Z$  is peaked at the boson mass, $m_{W}$ or $m_Z$.  The jets can be therefore tagged as $W$ and $Z$ by requiring $m_j$ to be close to $m_{W,Z}$. In the analysis of \cite{ATLAS_note}, a jet is identified as a $W$ if $m_j\in[69.4,95.4]$~GeV, and is identified as $Z$ if $m_j\in[79.8,105.8]$ GeV. The $W$ and $Z$ masses being close, these two ranges overlap. 
This implies there are  three disjoint regions to tag the jet:
\begin{itemize}
\item If $m_j\in[69.4,79.8]$, jet is $W$-only\,,
\item if $m_j\in[79.8,95.4]$, jet is $W$ or $Z$ (noted $W/Z$)\,,
\item if $m_j\in[95.4,105.8]$, jet is $Z$-only\,.
\end{itemize}
These tagging regions will be labelled by $I$. 
The expected $m_j$ distributions have been provided  in  Fig.~1c of \cite{ATLAS_note}. 
These distributions as well as the tagging regions are shown in Fig.~\ref{fig:mj_tag}. 

Note that the distributions for true $W$ and $Z$ have been generated assuming a bulk RS KK graviton signal. 
From Sec.~\ref{warped}, it is clear that $f_H^{-1}\gg f_V^{-1}$, so that the bulk RS KK graviton  decays mostly to longitudinally polarized $W$ and $Z$. However, the weak boson widths being narrow and the final shape being strongly widened by the detector effects, we expect the $W$, $Z$ distributions of Fig.~\ref{fig:mj_tag} to  hold for \textit{any polarisation} of the weak bosons to a very good approximation.

Using the distributions of Fig.~\ref{fig:mj_tag}, it is possible to estimate the tagging probabilities, given one of the two hypothesis for the  underlying true boosted particle, $\{{\rm True}\,\, $W$, {\rm True} \,\,$Z$\}$ that we will label by $X$. What we compute is thus the conditional probability $p(I|X)$. 
The conditional probabilities for tagging a true $W$ and a true $Z$ are computed from Fig.~\ref{fig:mj_tag} and  shown in Tab.~\ref{tab:CPjet}. These numbers are consistent with the ones found in \cite{Allanach:2015hba}. 
\begin{table}[h]
\center
\begin{tabular}{|c|ccc|}
\hline
$p(I|X)$ & $W$-only & $W$ or $Z$ & $Z$-only \\
\hline
True $W$ & 0.253 & 0.366 & 0.034 \\
\hline
True $Z$ & 0.112 & 0.398 & 0.211 \\ \hline
True $j$ & $0.025$ &  $0.035$ & $0.023$ \\ 
\hline
\end{tabular}
\caption{Conditional probabilities $p(I|X)$ for $W$, $Z$ and background jet tagging. \label{tab:CPjet}}
\end{table}
\begin{figure}
\center
\includegraphics[scale=0.3,clip=true, trim= 5.5cm 7.3cm 5cm 5cm]{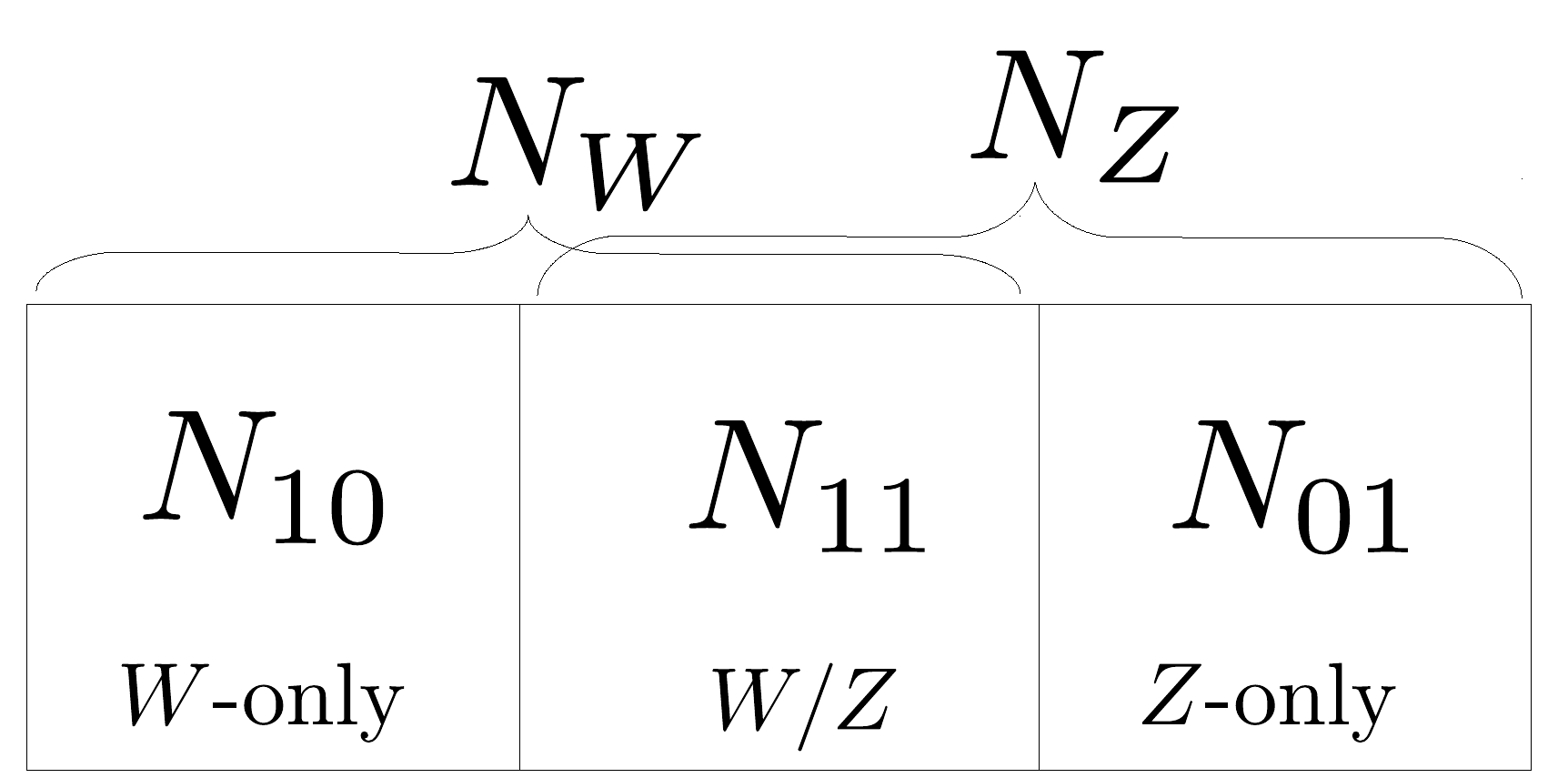}
\caption{ A picture summarizing the event numbers over the various weak-boson tagging regions. \label{fig:jet_regions}}
\end{figure}

Moreover, fat jets can also arise from the QCD interactions. 
The distribution for a jet coming from the QCD dijet background has been simulated in \cite{ATLAS_note} (see Figure~1\,c there), and appears to be nearly constant over the tagging regions. Using the simulated distributions, we can deduce the probabilities for mis-tagging a jet from the QCD background as a weak-boson jet.
Finally, the total probability for tagging a $W$, $Z$ or $j$ as a weak boson  $V$ 
 is just obtained by summing the probabilities over the three region. One gets $P(V|W)=65\%$, $P(V|Z)=72\%$, $P(V|j)=8\%$.

Before closing this subsection, it is instructive to focus on the counting statistics for the tagging of a single jet. This part can serve as a statistical toy-model for the upcoming analysis of the diboson excess. Indeed, most of the ingredients for the diboson analysis are already there, though applied to a simpler problem.

 Let us denote the tagging regions $W$-only, $W/Z$, $Z$-only as $10$, $11$, $01$, and labelled by $I\in{10,11,01}$. The first number of the region name means that the region potentially contains a $W$ if equal to one, and does not contain a $W$ if equal to zero. The second number of the name works similarly for the $Z$. These notations will be convenient later.
  
  The event numbers in each of these regions are denoted $N_{10}$ , $N_{11}$, $N_{01}$. These events  follow independent  Poisson statistics with parameter $\lambda_{10}$, $\lambda_{11}$, $\lambda_{01}$,
\be
P(N_{I}=n_I|\lambda_{I}) =\frac{\lambda_I^{n_I}e^{-\lambda_I}}{n_I!}\,. \label{eq:Pois}
\ee
Assuming an expected event number $\lambda_X=(\lambda_W,\lambda_Z)$ for the true $W$ and $Z$, the 
 $\lambda_I$ are expressed as 
 \be
 \lambda_I= \sum_X P(I|X) \lambda_X\,. \label{eq:lambdaIX}
 \ee
Equations~\eqref{eq:Pois},\eqref{eq:lambdaIX} put together provide $P(N_I=n_I|\lambda_X)$, the probability of observing $n_I$ events in the region $I$ for given expected event numbers $\lambda_X$.
Taking this probability as a function of $\lambda_X$ provides the likelihood function for $\lambda_X$, for an observed event number $n_I$.

Let us now assume that only the number of events that contain all possible $W$-tags and all possible $Z$-tags are reported. These numbers are defined as \be N_W=N_{10}+N_{11}\,,\quad N_Z=N_{01}+N_{11}\,.\ee This configuration is pictured in Fig.~\ref{fig:jet_regions}. 
Clearly, the statistics of $N_W$ and $N_Z$ are not independent, because of the common region $11$ where the jet is either $W$ or $Z$. Rather, the $N_W$, $N_Z$ follow a \textit{bivariate} Poisson statistics, given by 
\be
P(N_W=n_W,N_Z=n_Z|\lambda_I)=\sum_{\substack{n_{10}+n_{11}=n_W\\
n_{01}+n_{11}=n_Z}} \frac{\lambda_{01}^{n_{01}}\lambda_{10}^{n_{10}}\lambda_{11}^{n_{11}}}{n_{01}!n_{10}!n_{11}!}e^{-\lambda_{10}-\lambda_{01}-\lambda_{11}}
\label{eq:Pbi}
\ee
The mean of $(N_W,N_Z)$ is given by $(\lambda_{10}+\lambda_{11},\lambda_{01}+\lambda_{11})$, and the covariance matrix is
\be
\begin{pmatrix}
\lambda_{10}+\lambda_{11} & \lambda_{11} \\
\lambda_{11} & \lambda_{01}+\lambda_{11}
\end{pmatrix}\,.
\ee
 Plugging Eq.~\eqref{eq:lambdaIX} into Eq.~\eqref{eq:Pbi}, one gets the probability of getting $(n_W,n_Z)$ events for given expected event numbers $\lambda_X$.
Taking this probability as a function of $\lambda_X$ provides the likelihood function for $\lambda_X$, for an observed event number $n_W$ and $n_Z$.

\subsection{Statistics for the ATLAS diboson excess}

The probability for the tagging of two fat jets are obtained by combining the probability of tagging a single jet, see Tab.~\ref{tab:CPjet}. 
For the tagging of two jets, six tagging regions are obtained, by combining the labels $W$-only, $W/Z$ and $Z$-only in all inequivalent ways possible. The index $I$ of the tagging regions takes then the values 
\be
I\in\{(W,W),(W,Z),(Z,Z),(W,W/Z),(Z,W/Z),(W/Z,W/Z)\}\,.
\ee
The true events can be either a pair of weak bosons, a QCD jet mis-identified as a weak boson or two QCD jets mis-identified as weak bosons. The list of the hypothesis of true events, is then
\be
X\in\{(W,W),(W,Z),(Z,Z),(j,j) ,(W,j),(Z,j)\}\,,
\ee
where $j$ stands for background jet. The conditional probabilities $P(I|X)$ are given in Tab.\ref{tab:CPdijet}. The numbers for true $WW$, $WZ, ZZ$ are consistent with the ones reported in \cite{Allanach:2015hba}. 
 The dijet background corresponds to the true event $jj$.
Pileup effects  are assumed to be small, so that we do not consider the possibility of having true events as $(W,j)$, $(Z,j)$. On the other hand, one may consider a new physics signal giving rise to a $W$ and a jet or a $Z$ and a jet. We include therefore the probabilities $P(I|Wj_s)$, $P(I|Zj_s)$ in our table, assuming that the distribution of this signal jet $j_s$ is roughly the same as from a QCD jet. This case will not be considered in the rest of this work, as the decay of singlet resonances does not give rise to such signal.
\begin{table}
\resizebox{\linewidth}{!}{%
\begin{tabular}{|c|cccccc|}
\hline
$P(I|X)$ & $(W,W)$ & $(W,Z)$ & $(Z,Z)$ & $(W,W/Z)$ & $(Z,W/Z)$ & $(W/Z,W/Z)$\\ \hline\hline
 True $WW$ & 0.064 & 0.017 & 0.001 & 0.185  & 0.025 & 0.134 \\ \hline
 True $WZ$ & 0.028 & 0.057 & 0.007 & 0.142 & 0.091 & 0.146 \\ \hline
 True $ZZ$ & 0.013 & 0.047 & 0.045 & 0.089 & 0.168 & 0.158 \\ \hline
 True $jj$  & $6.25\cdot 10^{-4}$ & $11.5\cdot 10^{-4}$  & $5.29\cdot 10^{-4}$ & $17.5\cdot 10^{-4}$ & $16.1\cdot 10^{-4}$ & $12.3\cdot 10^{-4}$ \\ \hline\hline
 True $Wj_s$ & $6.33\cdot 10^{-3}$ & $6.7\cdot 10^{-3}$ & $0.78\cdot 10^{-3}$ & $18.0\cdot 10^{-3}$ & $9.61\cdot 10^{-3}$ & $12.8\cdot 10^{-3}$ \\ \hline
 True $Zj_s$ & $2.80\cdot 10^{-3}$ & $7.85\cdot 10^{-3}$ & $4.85\cdot 10^{-3}$ & $13.9\cdot 10^{-3}$ & $16.5\cdot 10^{-3}$ & $13.9\cdot 10^{-3}$\\ \hline
 \hline
\end{tabular}
}%
\caption{Conditional probabilities $p(I|{X})$ for tagging $WW$, $WZ$, $ZZ$ true events and a pair of background jets $jj$. The probabilities for tagging $Wj_s$ and $Zj_s$ are also included.
\label{tab:CPdijet}}
\end{table}

The number of events $N_I$  in each of the disjoint tagging regions $I$ follows a Poisson distribution with parameter $\lambda_I$, which is related to the expected number of true events (\ie~events before tagging) as 
\be
\lambda_I=\sum_X P(I|X) \lambda_X\,. \label{eq:PC_lambda}
\ee
The background expected event number $\lambda_{j j}$ will be obtained later on from the ATLAS analysis, once we know the statistics for the events.
The $\lambda_{WW}$, $\lambda_{WZ}$, $\lambda_{ZZ}$ are assumed to come only from the signal, \ie~the SM diboson background is neglected, following the ATLAS analysis. The $WW$, $WZ$, $ZZ$ expected event numbers are related to the total cross-sections by 
\be
\lambda_X = \epsilon_X B_X \mathcal{L}\, \sigma_X \,, \label{eq:lambdaX}
\ee
where $\mathcal{L}=20.3 $~fb$^{-1}$ is the integrated luminosity of the 2012 run,  $B_{WW}=B_W^2$, $B_{WZ}=B_WB_Z$, $B_{ZZ}=B_Z^2$  where $B_W=67.6\%$, $B_Z=69.9\%$ are the hadronic branching ratio of the weak bosons. 

The efficiencies $\epsilon'_X$ for selecting and tagging the signal are reported in \cite{ATLAS_note}, Fig.~2b. One gets roughly $\epsilon'_X\sim{0.10,0.13,0.09}$ with about $20\%$ of relative uncertainty. Note that these efficiencies are obtained assuming particular models.~\footnote{For example,  the bulk RS graviton used for the spin-2 simulation
and treated in Sec.~\ref{warped}  features the couplings $f_H^{-1}\gg f_V^{-1}$, so that it decays mostly to longitudinal polarisations. }
Slightly different efficiencies can be expected for different spins and  couplings. This model-dependence should be taken as an extra systematic uncertainty on the efficiencies.
As the weak-boson tagging probability based on the jet mass is already taken into account through the $P(I|X)$, it has to be removed from the $\epsilon_X'$ by dividing by $P(V|X)$. The efficiencies $\epsilon_X$ we will use are therefore given by 
\be
\epsilon_{WW}=\frac{\epsilon_{WW}'}{P(V|W)^2}\,,\quad \epsilon_{WZ}=\frac{\epsilon_{WZ}'}{P(V|W)P(V|Z)}\,\quad
\epsilon_{ZZ}=\frac{\epsilon_{ZZ}'}{P(V|Z)^2}\,,
\ee
so that $\epsilon_{X}\approx\{23\%, 28\%,  17\%\}$.

 In the ATLAS note \cite{ATLAS_note}, the expected event numbers $\lambda_I$ on the disjoint tagging regions are not reported. Rather, only the number of events that contain all possible $WW$-tags, $WZ$-tags and $ZZ$-tags are quoted. We denote them by $N_{WW}$, $N_{WZ}$, $N_{ZZ}$. It is convenient to label the tagging regions  with respect to their contribution to one or several of these reported rates. 
 
\begin{figure}
\center
\includegraphics[scale=0.3,clip=true, trim= 5.5cm 3cm 5cm 1cm]{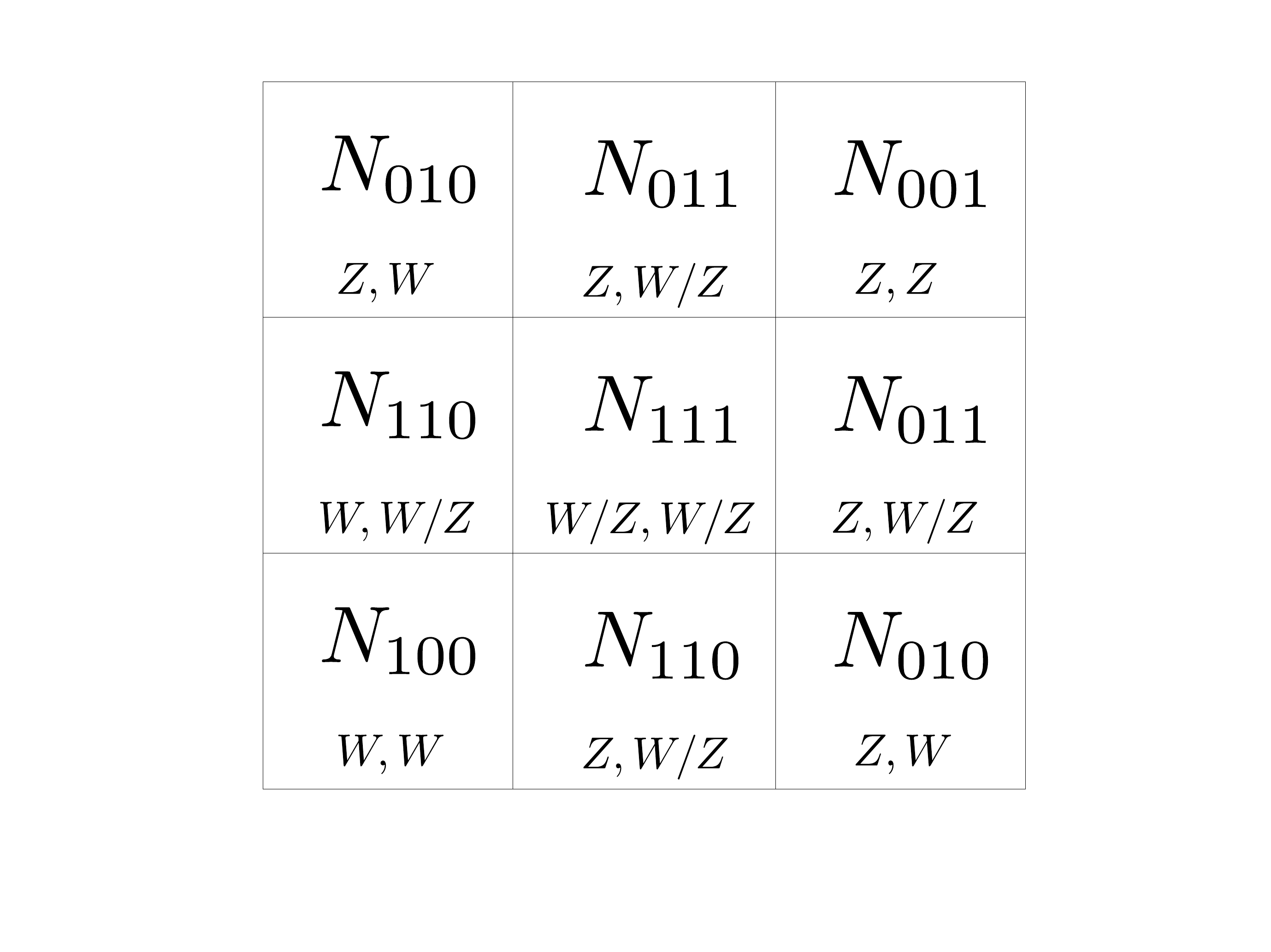}
\caption{ A picture summarizing the event numbers over the various diboson tagging regions. \label{fig:dijet_regions}}
\end{figure}
The labels are shown in Fig.~\ref{fig:jet_regions}. Using this parameterisation for the events, the observed events read
\begin{eqnarray}
N_{WW}&=&N_{100}+N_{110}+N_{111}\,,\nonumber\\
N_{WZ}&=&N_{010}+N_{110}+N_{011}+N_{111}\,,\\
N_{WW}&=&N_{001}+N_{011}+N_{111}\,,\nonumber
\end{eqnarray}
Clearly, these events are not independent. They rather follow a \textit{trivariate} Poisson statistics~\cite{kawamura1979},
\be
P(N_{WW}=n_{WW},N_{WZ}=n_{WZ},N_{ZZ}=n_{ZZ}|\lambda_I)= \sum_{{\cal D}}
\prod_{I}\frac{{\lambda_I}^{n_I}\,e^{-\lambda_I}}{n_I!}\label{eq:Ptri}
\,,
\ee
where the $n_I=\{n_{100}, n_{010}, n_{001}, n_{110}, n_{011}, n_{111}\}$ are positive integers running over the domain
\be\mathcal{D}=\mathcal{D}(n_{WW},  n_{WZ}, n_{ZZ})=\left\{n_I\left|\begin{array}{c}
n_{100}+n_{110}+n_{111}=n_{WW}\\
n_{010}+n_{110}+n_{011}+n_{111}=n_{WZ} \\
n_{001}+n_{011}+n_{111}=n_{ZZ}
\end{array}\right.\right\}\,.\label{eq:hatD}
\ee
The mean of this distribution is given by 
\begin{eqnarray}
\bar N_{WW}&=&\lambda_{100}+\lambda_{110}+\lambda_{111}\nonumber\\
\bar N_{WZ}&=&\lambda_{010}+\lambda_{110}+\lambda_{011}+\lambda_{111} \label{eq:mean_tri} \\
\bar N_{ZZ}&=&\lambda_{010}+\lambda_{011}+\lambda_{111}\nonumber 
\end{eqnarray}
The covariance matrix is given by 
\be
\begin{pmatrix}
\bar N_{WW} & \lambda_{110}+\lambda_{111} & \lambda_{111} \\
 & \bar N_{WZ} & \lambda_{011}+\lambda_{111} \\
 & & \bar N_{ZZ}
\end{pmatrix} \,.\label{eq:cov_tri}
\ee

The likelihood associated with the measured values of $\hat n_{WW}$, $\hat n_{WZ}$, $\hat n_{ZZ}$ is obtained by taking Eq.~\eqref{eq:Ptri} as a function of the hypothesis (\ie~$\lambda_I$) and using Eqs.~\eqref{eq:PC_lambda}. Dropping an irrelevant constant factor, the likelihood is a function of the the  various event numbers before tagging  $\lambda_X$ (recall that $X=\{WW,WZ,ZZ,jj\}$),
\be
L(\lambda_X)=\sum_{\hat{\mathcal{D}}} \prod_{I}\bigg(\sum_X P(I|X)\lambda_X  \bigg)^{n_I}e^{- \sum_X P(I|X) \lambda_X}\,,\label{eq:like_Pois}
\ee
where the observed event numbers appear through the domain $\hat{\mathcal{D}}\equiv{\mathcal{D}}(\hat n_{WW}, \hat n_{WZ}, \hat n_{ZZ}) $ and nowhere else. 
The $\lambda_{WW,\,WZ,\,ZZ}$ from the new physics signal are further related to the total production cross-sections by Eq.~\eqref{eq:lambdaX}. The evaluation of the  expected event number from dijet background $\lambda_{jj}$ is discussed in the next subsection.

\subsection{Consistency checks and the background likelihood}

As a consistency check of our  analysis, we can verify whether the event numbers tagged over the 
\textit{full} range $m_{jj}\in[1,3.5 ]$~TeV in the observed sample are consistent with the tagging rates determined in Tab.~\ref{tab:CPdijet} and with our statistical model leading to  Eq.~\eqref{eq:mean_tri}.
 The observed number of events  for the $WW$, $WZ$, $ZZ$ selections are given  in Tab.~8 of \cite{ATLAS_note}. These are $\hat n^{\rm full}_{WW}=425$, $\hat n^{\rm full}_{WZ}=604$, $\hat n^{\rm full}_{WZ}=333$.
  Regarding the expected rates, the complete region being overwhelmed by the dijet background, we can neglect the signal to a good approximation, so that the contributions to all tagging regions are simply proportional to the dijet expected event number over the full range, $\lambda_{j j}^{\rm full}$.
The ratios of the expected event numbers in the $WW$, $WZ$, $ZZ$ selections are then  obtained  using the tagging rates Tab.~\ref{tab:CPdijet}, Eq.~\eqref{eq:PC_lambda} and Eq.~\eqref{eq:mean_tri}.
It comes $n^{\rm full}_{WW}/n^{\rm full}_{WZ}=0.63$, $n^{\rm full}_{ZZ}/n^{\rm full}_{WZ}=0.57$
which are in agreement with the observed ratios  within $\sim 10\%$.

The statistical error on the ratios of the $\hat n^{\rm full}$ being roughly about $10\%$, this consistency check seems to be fulfilled within one standard deviation. However, this naive observation is too optimistic, because the event numbers of the  $WW$, $WZ$, $ZZ$ selections are actually strongly correlated. The joint statistics of the three selections is a trivariate Poisson, already described above, that now describes the whole  dataset (\ie~$m_{jj}\in[1,3.5 ]$~TeV). The mean and covariance matrix are thus given as in Eqs.~\eqref{eq:mean_tri}, \eqref{eq:cov_tri}.
 The covariance matrix reads 
\be
V_b^{\rm full}=\lambda_{j j}^{\rm full}\,\begin{pmatrix}
 3.60 & 2.98 & 1.23 \\
 2.98 & 5.74 & 2.84 \\
 1.23 & 2.84 & 3.37 
\end{pmatrix} \,,
\ee
where one used the values of $P(I|jj)$ obtained in Tab.~\ref{tab:CPdijet}.  The event numbers $\hat n^{\rm full}$ being large, one can adopt the Gaussian approximation so that the likelihood reads
  \be
L_b^{\rm full}(\lambda^{\rm full}_{jj})=\exp\bigg[-\frac{1}{2} \sum_{IJ} ( \hat n^{\rm full}_I-P(I|jj)\lambda^{\rm full}_{jj}) (V_b^{\rm full})_{IJ}^{-1}(\hat n^{\rm full}_J-P(J|jj)\lambda^{\rm full}_{jj})  \bigg]  \,.
  \ee
The maximum likelihood gives $-2\log L(\hat\lambda^{\rm full}_{jj})=10.6$. This value can readily be interpreted  as a compatibility test, whose statistics is a chi-square distribution with $3-1$ degrees of freedom. The equivalent statistical significance  obtained is $Z=2.6$. The compatibility is thus lower than the $1\sigma$ deviation naively found when neglecting correlations. This level of compatibility can nevertheless be considered as acceptable for high-energy physics standards, so that we pursue our analysis.

After these preliminary sanity checks, we now aim at building a consistent likelihood for the dijet background event number $\lambda_{jj}$ over the excess region $[1.85,2.15]$~TeV.  
The shape of the dijet background has been been estimated in \cite{ATLAS_note} using a smoothly falling distribution fitted to the observed dataset over the $m_{jj}\in[1,3.5 ]$~TeV range. A different fit is done for each of the three selections $WW$, $WZ$, $ZZ$. 
To the best of our understanding, each of these fits should give close results, because the only difference between the selections lies in the $m_j$ ranges selected. Comparing the $m_j$ intervals with the slope of the $m_{jj}$ shape, it appears that only  a slight decrease with $m_j$ of the efficiency of the boson-tagging cuts might be expected when going from the $W$-only to the  $Z$-only region. 

The outcome of the three fits can be seen in Tab.~6 and in Fig.~5 of \cite{ATLAS_note}. 
Comparing the central values obtained from the various fits using  the quoted error bars, it appears that  these  fits are compatible with each other only within roughly three standard deviations. Again, this naive comparison does not take into account the correlations, \ie~it assumes that the fits are independent from each other. These fits being partly based on the same dataset, their outcome are actually correlated, which implies that the actual uncertainty is smaller than the  one naively expected. This  implies that  the compatibility between the fits is worse than what naively expected.~\footnote{We point out that in the  current version of \cite{ATLAS_note}, the background model formula together with the best-fit values for the shape parameters do not reproduce at all the background curve shown in the plots. This presumed inconsistency  is another motivation to adopt a conservative approach for the background likelihood.} The shape systematic uncertainties evaluated in \cite{ATLAS_note} are found to be small, so that they cannot help solving this discrepancy.

In order to establish the dijet background likelihood using both a consistent and conservative approach, we shall \textit{(i)} take  the correlations among the fits into account and \textit{(ii)} assume somewhat larger  uncertainties than the ones quoted in \cite{ATLAS_note}. 
The likelihood for the expected dijet event number in the excess region $[1.85,2.15]$~TeV before tagging is approximately given by~\footnote{This is obtained by neglecting the shape systematic uncertainties and assuming standard error propagation. }
  \be
L_b(\lambda_{jj})=\exp\bigg[-\frac{1}{2} \sum_{I,J} ( b_I-P(I|jj)\lambda_{jj}) (V_{b})^{-1}_{IJ} ( b_J-P(J|jj)\lambda_{jj})  \bigg]  \,, \label{eq:likeb}
  \ee
  where the $P(I|jj)$ are given in Tab.~\ref{tab:CPdijet} and the expected values of $b_I$ obtained from the fits are given in Tab.~\ref{tab:data_bkg}. 
The covariance matrix $V_b$ is proportional to Eq.~\eqref{eq:cov_tri}, 
\be
V_b=\alpha\,\lambda_{jj}\,\begin{pmatrix}
 3.60 & 2.98 & 1.23 \\
 2.98 & 5.74 & 2.84 \\
 1.23 & 2.84 & 3.37 
\end{pmatrix} \cdot 10^{-3}
\ee
where $\alpha$ is a parameter that we tune to obtain a reasonable level of compatibility between the fits. As a criterion for the compatibility, we ask that $-2\log L_b(\hat\lambda_{jj})$ be equal to the level of compatibility obtained between the selections over $[1,3.5]$~TeV (see above). The criterion is thus \be-2\log L_b(\hat\lambda_{jj})\approx10.6\,.
\label{eq:crit}\ee

The maximum of the background likelihood $L_b(\lambda_{jj})$ is found to be
\be
\hat \lambda_{jj}=1422\,.
\ee
Besides, the coefficient satisfying the criterion Eq.~\eqref{eq:crit} is 
\be
\alpha\approx 0.377\,.
\ee
This fixes the overall amount of uncertainty, so that 
the background likelihood is fully determined.
The confidence intervals for $\lambda_{jj}$ are found to be 
\be
\lambda_{jj}\in[1151,1756]\textrm{ at 68\%CL}\,,\quad [935,2163]\textrm{ at 95\%CL}\,.
\ee
Note that the $68\%$ range translates as error bars $^{+1.20}_{-0.98}$, $^{+1.92}_{-1.55}$, $^{+1.13}_{-0.91}$ on the expected background events $b_{WW}$, $b_{WZ}$ and $b_{ZZ}$ respectively. As expected, these errors are larger than the ones quoted \citep{ATLAS_note}, that are shown in Tab.~\ref{tab:data_bkg}.

In order to model the systematic uncertainty on the background,
the likelihood $L_b$ will be included into the full likelihood, Eq.~\eqref{eq:fullL}.

\subsection{Analysis of total rates}

In the previous subsections, we have gradually derived the total likelihood that should be used to analyse the ATLAS diboson excess. It is given by the product of the likelihood derived from the counting statistics, Eq.~\eqref{eq:like_Pois}, times the likelihood constraining the background, given in Eq.~\eqref{eq:likeb}. In addition, as noted in \cite{Allanach:2015hba},  information on the counting of $WW+ZZ$ and $WW+WZ+ZZ$ are available in the additional material of \cite{ATLAS_note}. The values $\hat n_{WW+ZZ}=17$, $\hat n_{WW+WZ+ZZ}=17$ are reported. This introduces two new constraints on the event numbers $n_I$ of the disjoint tagging regions, 
\begin{eqnarray}
n_{100}+n_{001}+n_{110}+n_{011}+n_{111}=\hat n_{WW+ZZ} \\
n_{100}+n_{001}+n_{010}+n_{110}+n_{011}+n_{111}=\hat n_{WW+WZ+ZZ}\,,
\end{eqnarray}
that have to be added to the previous constraints already contained in $\hat{\mathcal{D}}$, see Eq.~\eqref{eq:hatD}.
It turns out that only three combinations are allowed, so that the domain $\hat{\mathcal{D}}$ is given by
\be
\hat{\mathcal{D}}'=\left\{ \begin{pmatrix}
n_{010}\\
n_{111}
\end{pmatrix}=
\begin{pmatrix}
0\\
5 \end{pmatrix},
\begin{pmatrix}
n_{100}\\
n_{001}\\
n_{110}\\
n_{011}\\
\end{pmatrix}=
\begin{pmatrix} 
2\\
0 \\
6 \\
4 
\end{pmatrix},
\begin{pmatrix}
1\\
1 \\
7 \\
3 
\end{pmatrix},
\begin{pmatrix}
0\\
2 \\
8 \\
2 
\end{pmatrix}
\right\}\,.
\ee
These numbers agree with the ones reported in  version~3 of \cite{Allanach:2015hba}. 
The final likelihood that we shall use to constrain the cross-sections for a hypothetical signal $\sigma_{Y}$, with $Y=\{WW,WZ,ZZ\}$  are thus
\be
\begin{split}
L(\sigma_Y)=&\int d\lambda_{jj}L_b(\lambda_{jj})\\ &\sum_{\hat{\mathcal{D}}'} \prod_{I}\bigg(\sum_Y P(I|Y)\epsilon_Y B_Y\sigma_Y +P(I|jj)\lambda_{jj} \bigg)^{n_I}e^{- \sum_Y P(I|Y)\epsilon_Y B_Y\sigma_Y+P(I|jj)\lambda_{jj}}\,.
\end{split}
\label{eq:fullL}
\ee
The  expected event number for the dijet background $\lambda_{jj}$ is constrained by $L_b(\lambda_{jj})$ and is treated as a nuisance parameter. The term in the second row agrees exactly with the likelihood used in version~3 of \cite{Allanach:2015hba}.

\begin{figure}[t]
\centering{
\includegraphics[scale=0.45]{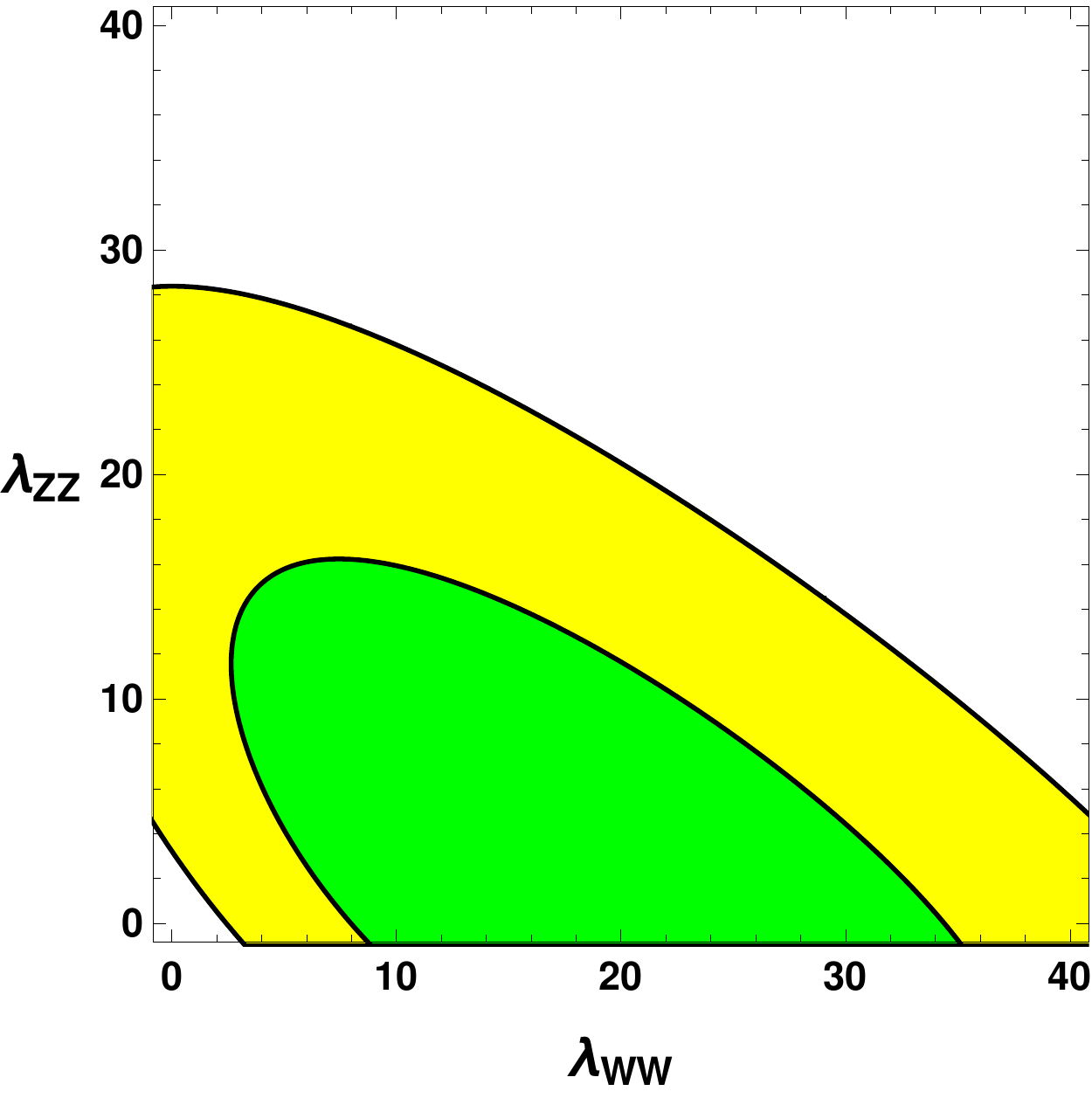}
\qquad
\includegraphics[scale=0.45]{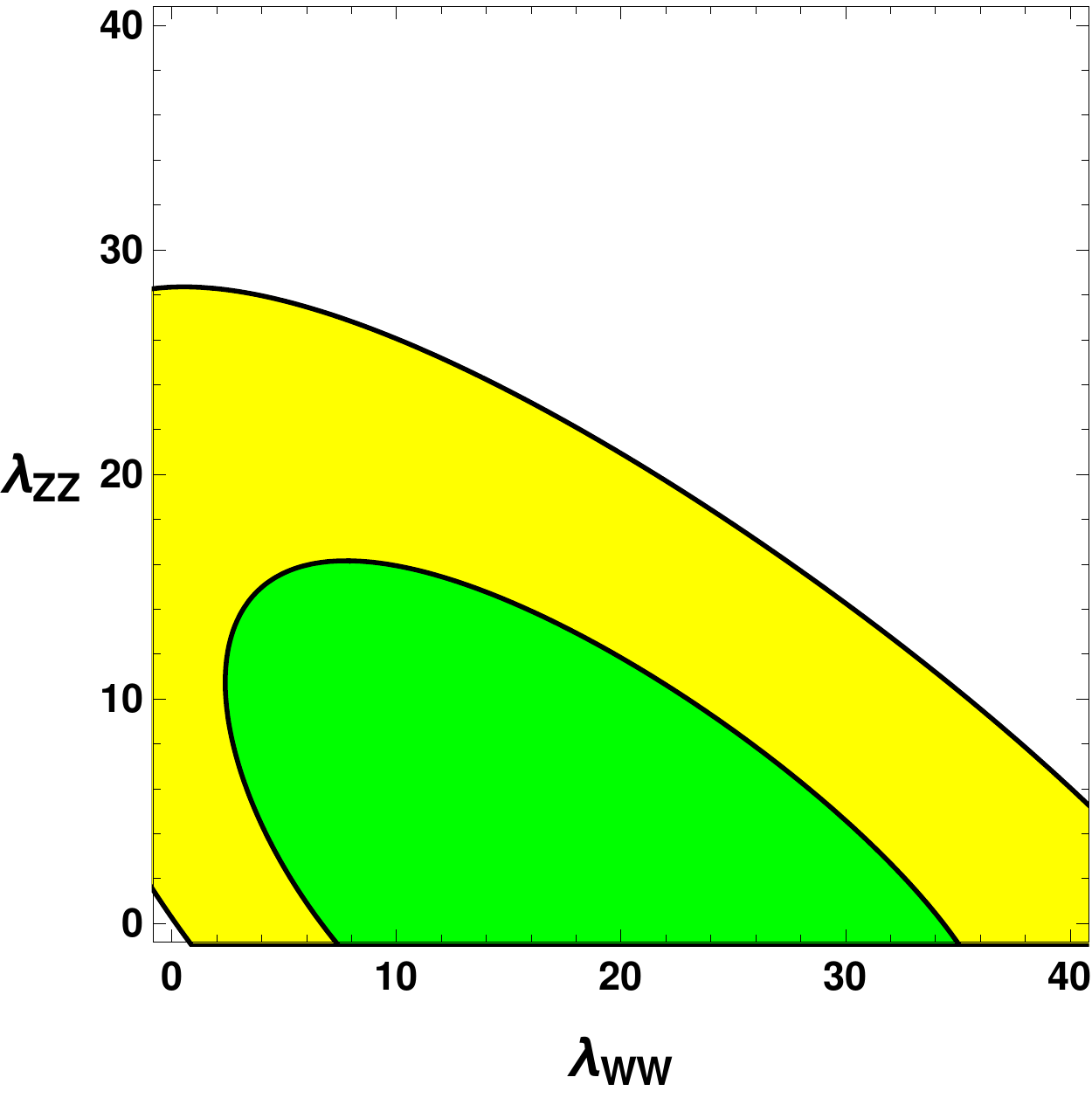}
}
\caption{ Posterior densities for the $WW$, $ZZ$ expected event numbers before tagging. The systematic uncertainty from the background determination is included in the right pannel. 
The $\lambda_{WZ}$ expected event number is set to zero. The $68\%$, $95\%$~C.L. regions are shown respectively in 
 green and yellow.
\label{fig:lambda2D}}
\end{figure}

Our interest being in neutral resonances, one should first compare the $H(\lambda_{WZ}=0)=\{\lambda_{WW}\neq0,\,\lambda_{ZZ}\neq0,\,\lambda_{WZ}=0\}$ hypothesis with the general hypothesis $H=\{\lambda_{WW}\neq0,\,\lambda_{ZZ}\neq0,\,\lambda_{WZ}\neq0\}$.  A consistent way to carry out such hypothesis testing is to  compute the Bayes factor
\be
B(\lambda_{WZ}=0)=\frac{p(H(\lambda_{WZ}=0)|{\rm data}))}{p(H|{\rm data})}=\left[\frac{\int d \lambda_{WW,ZZ} \, \pi(\lambda_{Y}) L(\lambda_{Y}) 
 }{\int d \lambda_{Y}\, \pi(\lambda_{Y}) L(\lambda_{Y})}
\frac{1}{\pi(\lambda_{WZ})} \right]_{\lambda_{WZ}=0}
 \,.
\ee
For the prior of the $\lambda_{Y}$, as described in Eq.~\eqref{eq:piPois}, we use 
 Poisson distributions with the Poisson parameter identified as the  expected number of background events $b_Y$, \ie~$P(b_Y+\lambda_Y|\lambda_Y)$. These priors arise from physical considerations and are conservative as they favour the background-only hypothesis.
We find
\be
B(\lambda_{WZ}=0)=0.96\,,
\ee
which implies that the $\lambda_{WZ}=0$ hypothesis is essentially as credible as the $\lambda_{WZ}\neq0$ hypothesis. This conclusion remains true for the $\lambda_{WZ}=0$ and $\lambda_{ZZ}=0$ hypothesis as well. On the other hand, the hypothesis with only $\lambda_{WZ}$ non-zero is highly disfavoured, with a Bayes factor  of $~2\cdot 10^{-5}$.

We  then proceed by drawing the best-fit regions for $\sigma_{WW}$, $\sigma_{ZZ}$ from the posterior $L(\sigma_{WW,ZZ})\pi(\sigma_{WW,ZZ})$. The priors for the cross sections are taken  flat. If one does not taken into the uncertainty on the background, the regions obtained are shown in the left pannel of Fig.~\ref{fig:lambda2D}. 
Finally, these regions of $\lambda_{WW,ZZ}$ are readily translated into regions for total cross sections $\sigma_{WW}$, $\sigma_{ZZ}$, that are shown in Fig.~\ref{fig:lambda2D}.

\section{Interpreting the diboson excess with the neutral resonances EFT}
\label{se:results}

\begin{figure}[t]
\centering{
\includegraphics[scale=0.55]{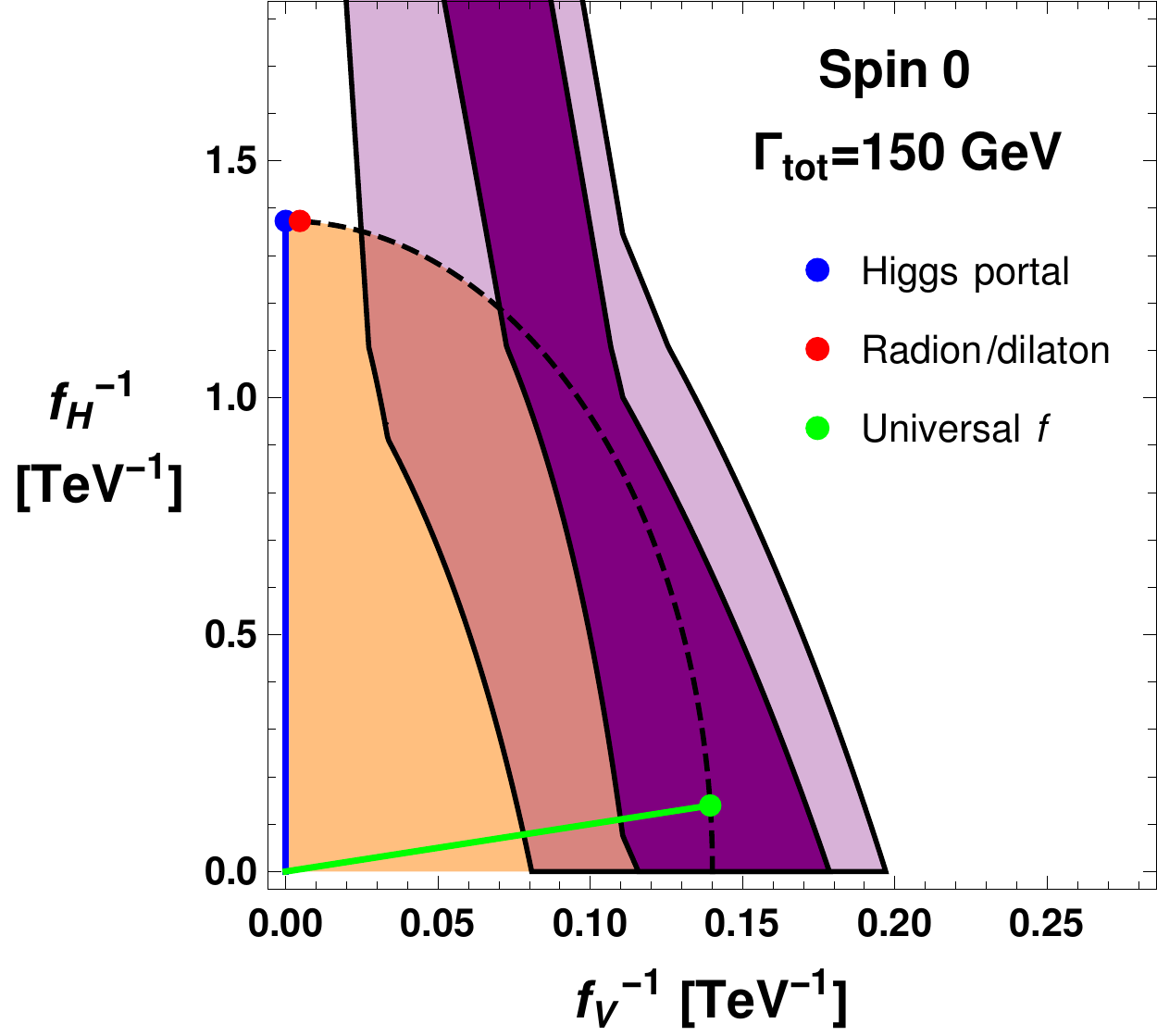}
\qquad
\includegraphics[scale=0.55]{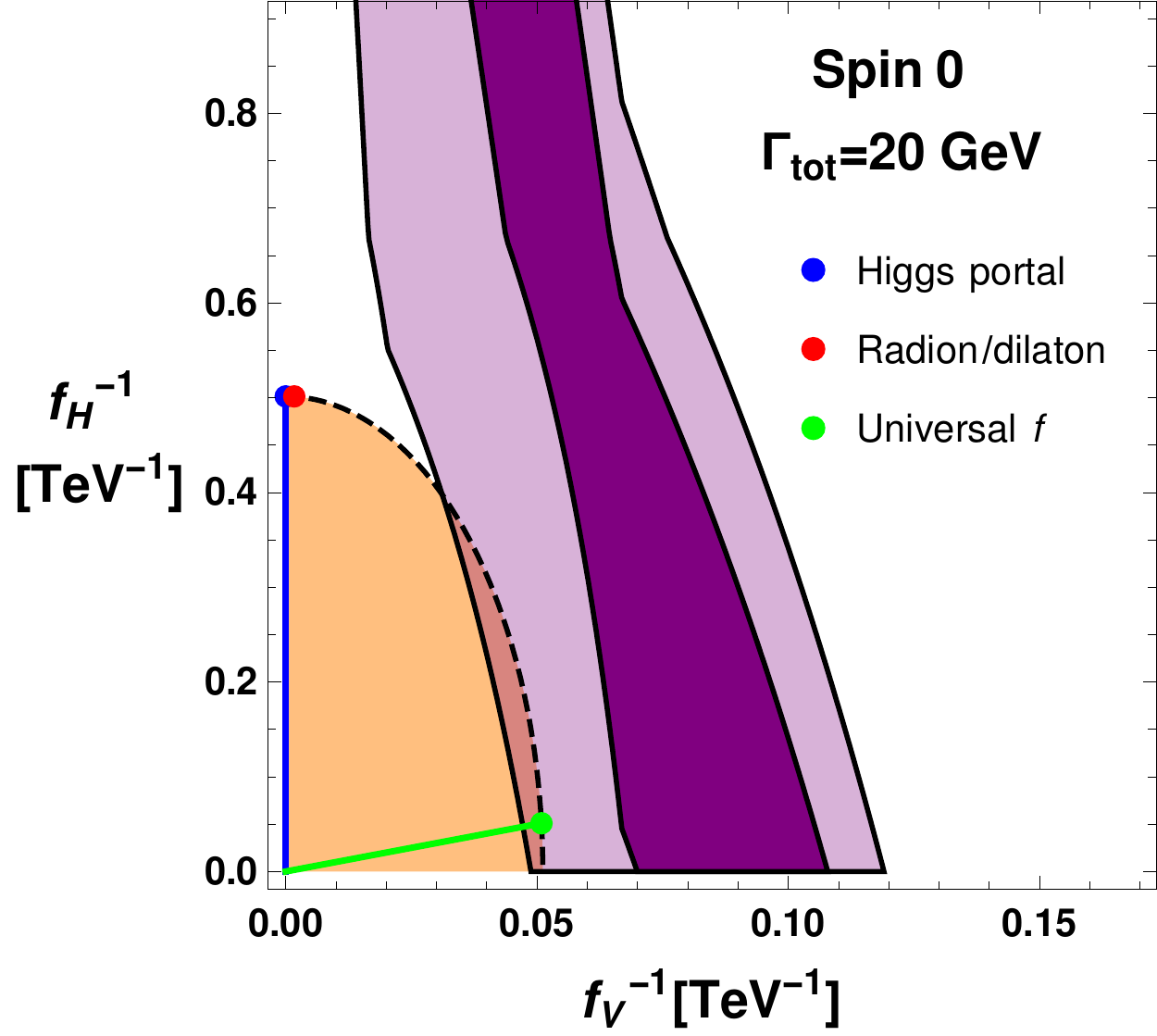}
}

\caption{ Constraints from the ATLAS diboson excess \cite{ATLAS_note} on a generic neutral resonance with spin 0. A total width  of $150$~GeV and $20$~GeV is assumed on the left and right pannel respectively. 
Purple regions correspond to $68\%$, $95\%$ CL regions drawn from the total event number in the excess region. The orange region is the one potentially  compatible with the assumed total width.
This bound is saturated when the partial widths from $\mathcal{O}_V$, $\mathcal{O}_H$ are the only contribution to the total width.
\label{fig:fVfHspin0}}
\end{figure}

In the new physics scenarios considered in Sec.~\ref{se:scenarios},  the neutral resonance couplings to field strengths are universal. We will therefore make a simplifying assumptions and use a single parameter
\be
f_V \equiv f_W=f_B=f_G\,,
\ee
both in the spin-0 and spin-2 cases.
We then focus on the tree-level production induced by the $\mathcal{O}_H$, $\mathcal{O}_V$ operators via gluon fusion (GGF) and weak boson fusion (VBF). We find that VBF is subleading to GGF for most of the parameter space.
For the spin-0 case, these two operators also completely fix the width $\Gamma_{\rm SM}$ into SM particles (up to suppressed contributions from $\mathcal O_T$), while for 
 the spin-2 case, one can have contributions from the operators $\mathcal O_\psi$ if present.~\footnote{In our analysis we do not take into account 
 production of the spin-2 resonance via quark-fusion  (which can be induced by the operators $\mathcal O_\psi$ for $\psi=q_L, u_R,d_R$), nor NLO-QCD effects. 
Both effects have recently been considered in Ref.~\cite{Li:2014awa} and were shown to lead to  $\mathcal O(1)$ modifications of the production rates.
} In addition one can allow for an invisible width, \ie~$\Gamma_{\rm tot}=\Gamma_{\rm SM}+\Gamma_{\rm inv}$ into non-SM particles.
The total width estimation from the shape analysis of Sec.~\ref{se:stat_shape} then provides us with another constraint in the $f_V-f_H$ plane, \ie~  
 \be \Gamma_{V}+\Gamma_{H}\label{eq:widthregion}\leq \Gamma_{\rm tot}\,,\ee 
 where the  $\Gamma_{H,V}$ are the partial widths induced by the $\mathcal{O}_H$, $\mathcal{O}_V$ operators.

We write  {\tt FeynRules}~\cite{FR} models for  the EFT of neutral resonances described in  Sec.~\ref{se:eft}.
The signal expected from the spin 0 and 2 resonances  $pp\rightarrow \phi \rightarrow  WW,ZZ$, and $pp\rightarrow \phi_{\mu\nu} \rightarrow  WW,ZZ$ at the 8 TeV LHC  are then computed using  {\tt MadGraph 5}~\cite{MG}.  The main cuts  are $p_T>540$ GeV and $|\eta|<2$ for each of the outgoing vector bosons, which we implement using {\tt MadAnalysis 5}~\cite{MA}.

The limits on the spin-0 CP even resonance are shown in Fig.~\ref{fig:fVfHspin0}. We choose  $\Gamma_{\rm tot}$ within the 95\% confidence interval provided in Tab.~\ref{tab:1DBCI}, \ie~we fix it to $\Gamma_{\rm tot}=150$ GeV (left panel of Fig.~\ref{fig:fVfHspin0}) and 
$\Gamma_{\rm tot}=20$ GeV (right panel of Fig.~\ref{fig:fVfHspin0}). The orange shaded regions are those allowed by the condition \eqref{eq:widthregion}.
We also display how various scenarios fit into the shown parameter space. The Higgs portal scenario, which is indistinguishable from the radion in the RS brane model, is shown as the blue point. The line emerging from that point would correspond to a hypothetical model where the scalar boson can decay into invisible states. 
Similarly, the radion of the RS bulk model is depicted as the red point. Neither scenario can fit the observed excess, mainly because the operator $\mathcal O_G $ is not available for GGF production. Generating sufficient contribution from the VBF process  would require too small values of $f_H$, in conflict with the measured width.
Finally, the green line shows a hypothetical  scalar with universal $f_H=f_V\equiv f$. This scenario could explain the required width and production rate for $f=7$ TeV (19 TeV)  for $\Gamma_{\rm tot}=150$ GeV (20 GeV).

\begin{figure}
\begin{picture}(400,200)
\put(0,0){\includegraphics[scale=0.55]{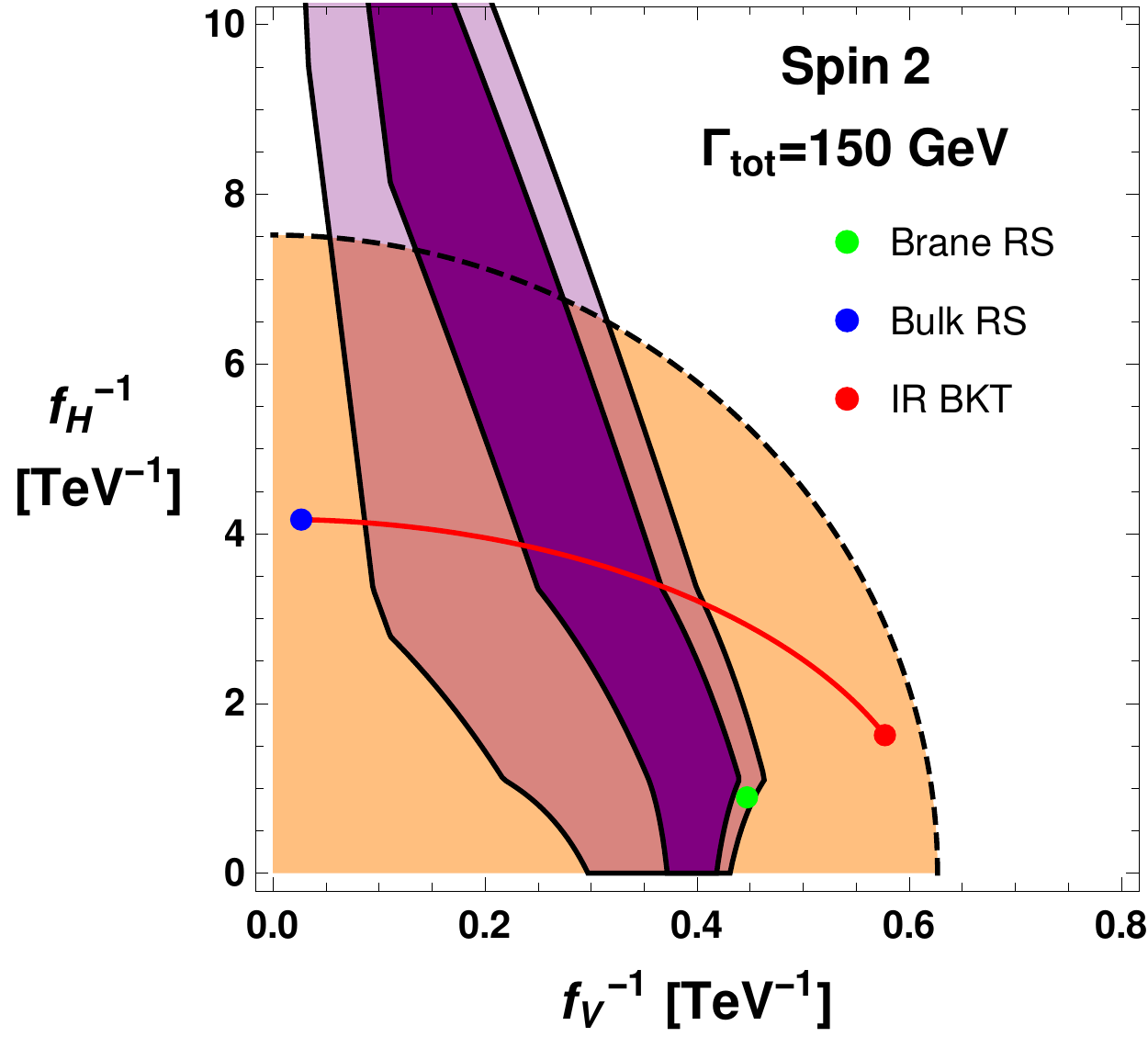}}
\put(220,0){\includegraphics[scale=0.55]{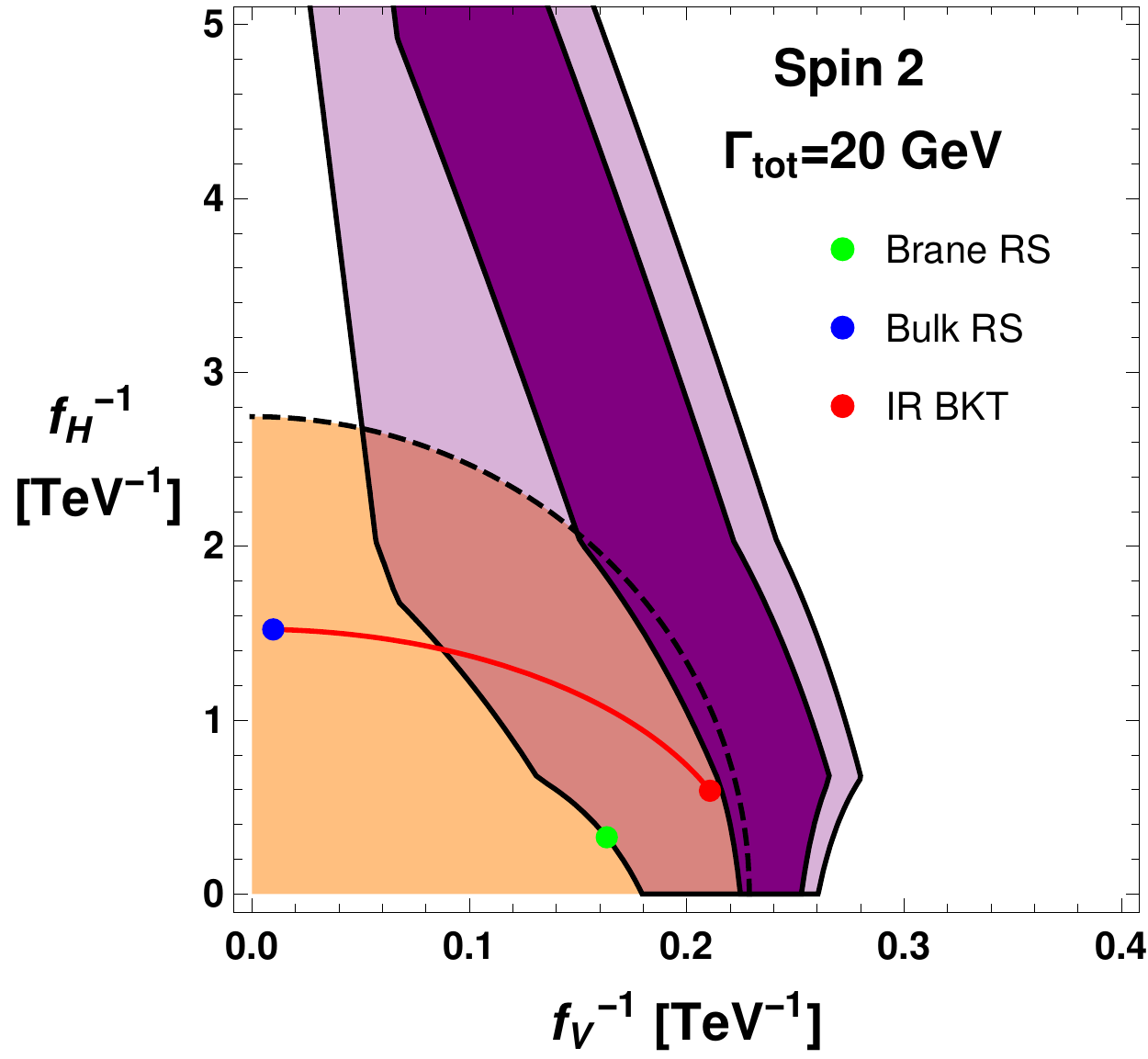}}
\end{picture}
\caption{ Constraints from the ATLAS diboson excess \cite{ATLAS_note} on a generic neutral resonance with spin 2. A total width  of $150$~GeV and $20$~GeV is assumed on the left and right pannel respectively. 
Purple regions correspond to $68\%$, $95\%$ CL regions drawn from the total event number in the excess region. The orange region is the zone potentially  compatible with the total width assumed.
This bound is saturated when the partial widths from $\mathcal{O}_V$, $\mathcal{O}_H$ are the only contribution to the total width.
\label{fig:fVfHspin2}}
\end{figure}

We present the analogous limits on spin-2 resonances in Fig.~\ref{fig:fVfHspin2}. The RS brane model is shown as the green point corresponding to the value of $\kappa$ obtained from the chosen width of the resonance, again 150 GeV (left) and 20 GeV (right). The implicit values for the coupling are $\kappa=0.23$ (for $\Gamma_{\rm tot}=150$ GeV) and $\kappa=0.09$ (for $\Gamma_{\rm tot}=20$ GeV).
Note that $\Gamma_H+\Gamma_V\approx 0.52\, \Gamma_{\rm tot}$, the rest being contributed by the fermions.
The RS bulk model 
with universal brane kinetic terms 
for the gauge fields has two parameters ($r_1$ and $\kappa$), and hence for fixed width it corresponds to one-dimensional curves in the parameter space, shown as the red curves. We vary $0<r_1<90$ (see footnote \ref{foot:pert}), corresponding to implicit values of $\kappa =1.09-0.42$ (for $\Gamma_{\rm tot}=150$ GeV) and 
$\kappa=0.40-0.16$ (for $\Gamma_{\rm tot}=20$ GeV).
One can see that the brane model can in fact explain the observed excess, while the canonical RS bulk model (with $r_1=0$, the blue point) cannot, because the values of  $\kappa$ needed  to fit the correct width and production cross section are in conflict with each other. On the other hand, allowing for IR BKT's, one can fit both width and total production rate of the excess.
The required size of the BKT's is only $r_1\sim 1-4$, depending on $\Gamma_{\rm tot}$.

\section{Conclusion}

New particles, singlets under the SM interactions and with masses near the TeV scale can arise in many well-motivated extensions of the SM, including extra dimensions, strongly-coupled scenarios as well as the Higgs portal. Such particles can be linearly coupled to SM operators, and can thus appear as resonances in $s$-channel processes. 

In this paper we first lay down the complete effective Lagrangians for neutral resonances of spin 0 and 2. It turns out that this EFT consists of only  few operators, which can further be restricted by theoretically well-motivated assumptions, such as approximate flavor and CP conservation. 
Given the concise description of a large class of models in terms of few parameters, our EFT can serve as a model-independent framework to study the implications for any resonance searches at the LHC. We compute the generic widths of the resonances and present explicitly the matching to the  new physics scenarios quoted above.


We then investigate the possibility that a new heavy resonance be at the origin of the excess in diboson production recently reported by the ATLAS collaboration.
We compute the local  significances in both frequentist and Bayesian frameworks, showing a moderate evidence for the existence of a signal. We perform a shape analysis of the excess under full consideration of the systematic uncertainties to extract the width $\Gamma_{\rm tot}$ of the hypothetical resonance, finding it to be in the range 26 GeV $<\Gamma_{\rm tot}<$ 144 GeV at 95\% C.L. 
Turning to the study of total event numbers, we first evaluate the conditional probabilities for tagging  a true $W$, $Z$ and QCD jet as either $W$ or $Z$  from the ATLAS simulations. From these one deduces  the tagging probabilities for the $WW$, $WZ$, $ZZ$ selections reported by ATLAS. 
 We further observe that these three overlapping selections follow a joint trivariate Poisson distribution, which opens the possibility of a thorough likelihood analysis of the event rates.
The tagging probabilities are checked against  the full observed sample.
A conservative treatment  of the dijet background is adopted, that includes  the correlations among the three selections $WW$, $WZ$, $ZZ$. The uncertainty on this background estimation  is then  taken into account as a systematic error in the total likelihood. Finally, using an actual hypothesis testing, we show that the data do not require $WZ$ production and can thus in principle be explained by neutral resonances.

Finally, we test the effective Lagrangians of neutral resonance using both the information from the width and the cross section of the analysis of the ATLAS data. It turns out that these pieces of information imply stringent contraints on the EFT parameter space once put together, even after including the background uncertainty. These exclusion bounds further imply that various popular scenarios appear to be totally incompatible with the ATLAS diboson excess.
We find that neither scalars coupled via the Higgs-portal nor the RS radion can explain the ATLAS anomaly. The RS  graviton with all matter on the IR brane can in principle fit the observed excess, while the RS model with matter propagating in the bulk requires the presence of IR brane kinentic terms for the gauge fields.

As an outlook, we emphasize that it would be interesting to constrain the EFT for neutral resonances using other LHC searches. As the effective Lagrangians are rather predictive, powerful conclusions can be expected by combining the information from various channels.


\section*{Acknowledgements} 

We would like to thank Benjamin Fuks, Olivier Mattelaer and Claude Dhur for their help with the automated tools, and the conveners of the ATLAS exotics group for clarifications regarding \cite{ATLAS_note}.
We acknowledge the Funda\c c\~ao de Amparo \`a Pesquisa do Estado de S\~ao Paulo (FAPESP) for financial support.

\appendix

\section{Polarization tensors for spin-2 fields.}
\label{polarizations}

The polarization tensors for massive spin-2 fields satisfy 
\be
k^\mu\epsilon_{\mu\nu}^s=0,\,,\qquad \epsilon^{s\,\mu}_{\ \ \mu}=0\,,\qquad \epsilon^s_{\mu\nu}=\epsilon^s_{\nu\mu}
\ee
and their orthogonality and completeness relations read \cite{Hinterbichler:2011tt}
\be
\epsilon_{\mu\nu}^{*s}\epsilon^{s\,\rho\sigma}=\Pi^{\rho \sigma}_{\mu\nu}\,,\qquad \epsilon^{*s}_{\mu\nu}\epsilon^{s'\,\mu\nu}=\delta^{ss'}
\ee
where the projector $\Pi$ was given in Eq.~(\ref{projector}).
They can be written in terms of tensor products of spin-1 polarization vectors
\be
\epsilon^s_{\mu\nu}=
\left\{ \epsilon^+_\mu\epsilon^+_\nu\,,\, \frac{1}{\sqrt{2}}\, (\epsilon^0_{\mu}\epsilon^+_{\nu}+\epsilon^+_{\mu}\epsilon^0_{\nu})\,,\,
\frac{1}{\sqrt{6}}
(\epsilon^+_{\mu}\epsilon^-_{\nu}+\epsilon^-_{\mu}\epsilon^+_{\nu}-2\epsilon^0_\mu\epsilon^0_\nu)\,,\,
\frac{1}{\sqrt{2}}\, (\epsilon^0_{\mu}\epsilon^-_{\nu}+\epsilon^-_{\mu}\epsilon^0_{\nu})\,,\,
\epsilon^-_\mu\epsilon^-_\nu
\right\}\,.
\ee
We note that the fields $\tilde h_{\mu\nu}^n$ in Ref.~\cite{Han:1998sg} are not canonically normalized, and correspondingly their polarization tensors (propagators) are $\sqrt{2}$ (2) times ours.

\section{Partial widths}
\label{app:partial}

In Sec.~\ref{se:eft} we gave the total widths of the various resonances in function of the effective couplings. The partial widths, if required, can easily be obtained from these formulae. For the field-strength couplings, one can use the decomposition
\be
\frac{3}{f_W^2}+\frac{1}{f_B^2}=\frac{2}{f_W^2}+\left(\frac{c_w^2}{f_W}+\frac{s_w^2}{f_B}\right)^2
+\left(\frac{s_w^2}{f_W}+\frac{c_w^2}{f_B}\right)^2+2c_w^2s_w^2\left(\frac{1}{f_W}-\frac{1}{f_B}\right)^2
\ee
where the four terms correspond to $WW$, $ZZ$, $\gamma\gamma$ and $Z\gamma$ decays respectively
For the coupling $f_H$ one has simply $f_H^{-2}=\frac{1}{2}f_H^{-2}+\frac{1}{4}f_H^{-2}+\frac{1}{4}f_H^{-2}$, corresponding to (longitudinal) $WW$, $ZZ$, and $hh$ decays. Finally, for the partial widths of the fermions one can use decompositions such as
\be
\frac{N_{q_L} }{f_{q_L}}+\frac{N_{t_R}}{f_{t_R}}+\frac{N_{b_R}}{f_{b_R}}=3\left(\frac{1}{f_{q_L}}+\frac{1}{f_{t_R}}\right)+3\left(\frac{1}{f_{q_L}}+\frac{1}{f_{b_R}}\right)
\ee
corresponding to the partial widths of the top and bottom quarks, and similarly for the other SM fermions.


\noindent
\clearpage

\bibliographystyle{JHEP} 

\bibliography{ATLAS_diboson_final}

\end{document}